\documentclass[aps,prl,twocolumn,reprint,amsmath,amssymb,superscriptaddress,floatfix,footinbib,longbibliography]{revtex4-2}

\usepackage{graphicx}
\usepackage{epstopdf}

\usepackage[T1]{fontenc}
\usepackage[applemac]{inputenc}
\usepackage{lmodern}
\usepackage{amsmath}
\usepackage{amssymb}
\usepackage[english]{babel}
\usepackage{natbib}
\usepackage{ae}
\usepackage{units}
\usepackage{upgreek}

\usepackage{amsmath,amssymb,natbib,bm}
\usepackage{psfrag}
\usepackage{subfigure}
\usepackage{amsthm}

\usepackage{booktabs}

\usepackage{slashed}

\usepackage[americaninductors]{circuitikz}
\usepackage{tikz}
\usetikzlibrary{arrows}
\usepackage{ulem}
\usepackage{color}
\usepackage{url}

\usepackage[colorlinks]{hyperref}
\hypersetup{%
        plainpages=true,
        breaklinks=true,
        hypertexnames=false,
        pageanchor=true,
        colorlinks=true,
        linkcolor={blue},
        citecolor={magenta},
        urlcolor={blue},
        anchorcolor={black}
      }

\usepackage{mleftright} 

\newcommand{\figref}[1]{\mbox{Fig.~\ref{#1}}}

\renewcommand{\eqref}[1]{\mbox{Eq.~(\ref{#1})}}
\newcommand{\eqsref}[2]{\mbox{Eqs.~(\ref{#1})--(\ref{#2})}}
\newcommand{\eqsrefCross}[2]{\mbox{Eqs.~(\ref{#1}), (\ref{#2})}}

\newcommand{\figpanel}[2]{Fig.~\hyperref[#1]{\ref*{#1}(#2)}} 
\newcommand{\figpanels}[3]{Fig.~\hyperref[#1]{\ref*{#1}(#2)-(#3)}} 
\newcommand{\figpanelNoPrefix}[2]{\hyperref[#1]{\ref*{#1}(#2)}} 
\newcommand{\figpanelsNoPrefix}[3]{\hyperref[#1]{\ref*{#1}(#2)-(#3)}} 

\newcommand{\ket}[1]{|#1\rangle}

\newcommand{\phase}[1]{\mathrm{Arg}(#1)}

\newcommand{\be}{\begin{equation}}
\newcommand{\ee}{\end{equation}}
\newcommand{\bea}{\begin{eqnarray}}
\newcommand{\eea}{\end{eqnarray}}

\begin{document}

\title{Group delay controlled by the decoherence of a single artificial atom}

\author{Y.-T.~Cheng}
\thanks{These authors contributed equally}
\affiliation{Department of Physics, City University of Hong Kong, Kowloon, Hong Kong SAR 999077, China}

\author{K.-M.~Hsieh}
\thanks{These authors contributed equally}
\affiliation{Department of Physics, City University of Hong Kong, Kowloon, Hong Kong SAR 999077, China}

\author{B.-Y.~Wu}
\thanks{These authors contributed equally}
\affiliation{Department of Physics, City University of Hong Kong, Kowloon, Hong Kong SAR 999077, China}

\author{Z.~Q.~Niu}
\thanks{These authors contributed equally}
\affiliation{State Key Laboratory of Materials for Integrated Circuits, Shanghai Institute of Microsystem and Information Technology (SIMIT), Chinese Academy of Sciences, Shanghai 200050, China}
\affiliation{ShanghaiTech University, Shanghai 201210, China}

\author{F.~Aziz}
\affiliation{Department of Physics, National Tsing Hua University, Hsinchu 30013, Taiwan}

\author{Y.-H.~Huang}
\affiliation{Department of Physics, National Tsing Hua University, Hsinchu 30013, Taiwan}

\author{P.~Y.~Wen}
\affiliation{Department of Physics, National Chung Cheng University, Chiayi 621301, Taiwan}

\author{K.-T.~Lin}
\affiliation{CQSE, Department of Physics, National Taiwan University, Taipei 10617, Taiwan}

\author{Y.-H. Lin}
\affiliation{Department of Physics, National Tsing Hua University, Hsinchu 30013, Taiwan}
\affiliation{Center for Quantum Technology, National Tsing Hua University, Hsinchu 30013, Taiwan}

\author{J.~C.~Chen}
\affiliation{Department of Physics, National Tsing Hua University, Hsinchu 30013, Taiwan}
\affiliation{Center for Quantum Technology, National Tsing Hua University, Hsinchu 30013, Taiwan}

\author{A.~F.~Kockum}
\affiliation{Department of Microtechnology and Nanoscience (MC2), Chalmers University of Technology, SE-412 96 Gothenburg, Sweden}

\author{G.-D.~Lin}
\affiliation{CQSE, Department of Physics, National Taiwan University, Taipei 10617, Taiwan}
\affiliation{Physics Division, National Center for Theoretical Sciences, Taipei 10617, Taiwan}

\author{Z.-R.~Lin}
\email[e-mail:]{zrlin@mail.sim.ac.cn}
\affiliation{State Key Laboratory of Materials for Integrated Circuits, Shanghai Institute of Microsystem and Information Technology (SIMIT), Chinese Academy of Sciences, Shanghai 200050, China}

\author{Y.~Lu}
\email[e-mail:]{kdluyong@outlook.com}
\affiliation{Guangzhou Institute of technology, Xidian University, Xi'an, China}
\affiliation{Advanced Interdisciplinary Research Center, Xidian University, Xi'an, China}
\affiliation{Faculty of Integrated Circuit, Xidian University, Xi'an, China}

\author{I.-C.~Hoi}
\email[e-mail:]{iochoi@cityu.edu.hk}
\affiliation{Department of Physics, City University of Hong Kong, Kowloon, Hong Kong SAR 999077, China}

\date{\today}

\begin{abstract}

The ability to slow down light at the single-photon level has applications in quantum information processing and other quantum technologies.
We demonstrate two methods, both using just a single artificial atom, enabling dynamic control over microwave light velocities in waveguide quantum electrodynamics (waveguide QED).
Our methods are based on two distinct mechanisms harnessing the balance between radiative and non-radiative decay rates of a superconducting artificial atom in front of a mirror.
In the first method, we tune the radiative decay of the atom using interference effects due to the mirror; in the second method, we pump the atom to control its non-radiative decay through the Autler--Townes effect.
When the half the radiative decay rate exceeds the non-radiative decay rate, we observe positive group delay; conversely, dominance of the non-radiative decay rate results in negative group delay.
Our results advance signal-processing capabilities in waveguide QED.

\end{abstract}

\maketitle


\textit{Introduction.}---Slow light has attracted great attention in recent years with impressive progress such as slowing down~\cite{Hau1999, Khurgin2010} and even stopping light completely~\cite{Phillips2001}.
This progress opens up for potential applications in quantum information technology, e.g., quantum memories or synchronizing qubits for parallel computing~\cite{Lvovsky2009}.
The benefits of slow and fast light extend to other applications as well, including enhancing nonlinear interactions~\cite{Kash1999, Hamachi2009, Monat2009, Li2011}, pulse compression~\cite{Kondo2015, Ossiander2021}, and increasing sensitivity in interferometers~\cite{Shi2007, Shahriar2007} and gas sensing~\cite{Lai2011}.
Typically, slow light is achieved through electromagnetically induced transparency (EIT)~\cite{Kasapi1995, Budker1999, He2007, Shiau2011, Novikov2016, Rastogi2019} when light interacts with multi-level atomic systems.
In such setups, EIT provides both a steep dispersion profile, directly related to the speed of the slowed light, and high transmission.
On the other hand, fast light has been observed in atomic gases~\cite{Wang2000, Dogariu2001}, in crystals~\cite{Bigelow2003, Manipatruni2008}, and in rubidium vapor~\cite{Mirhosseini2016}.
To enable flexible manipulation of light signals, the capability to switch between slow and fast light in situ is crucial~\cite{Zhao2021, Lu2022, Akram2015, Brehm2022, Zheng2023}, especially for signal processing in waveguide QED at single-photon level~\cite{vanLoo2013, Lalumi2013, Koshino2012, Gheeraert2020, Kannan2023, Kannan2024, You2011, You2003, Zhou2008}.

In this Letter, we demonstrate that the speed of light can be controlled by the decoherence of a single atom.
Specifically, by changing the ratio of the radiative and non-radiative decay rates ($\Gamma_{10}$ and $\Gamma_{10}^n$, respectively) of a two-level artificial atom ($\ket{0} \leftrightarrow \ket{1}$ transition of a superconducting transmon qubit~\cite{Koch2007}) coupling to light near the end of a waveguide [a one-dimensional transmission line (TL)], we can control the group delay time $\tau_d$ of the light, which affects the group velocity $v_g$ of the light.
The underlying mechanism for the control of $\tau_d$ is that these decay rates, together with the detuning between the light and the atomic transition, set the reflection coefficient $r$ of the atom--mirror system.

\begin{figure}
\includegraphics[width=\linewidth]{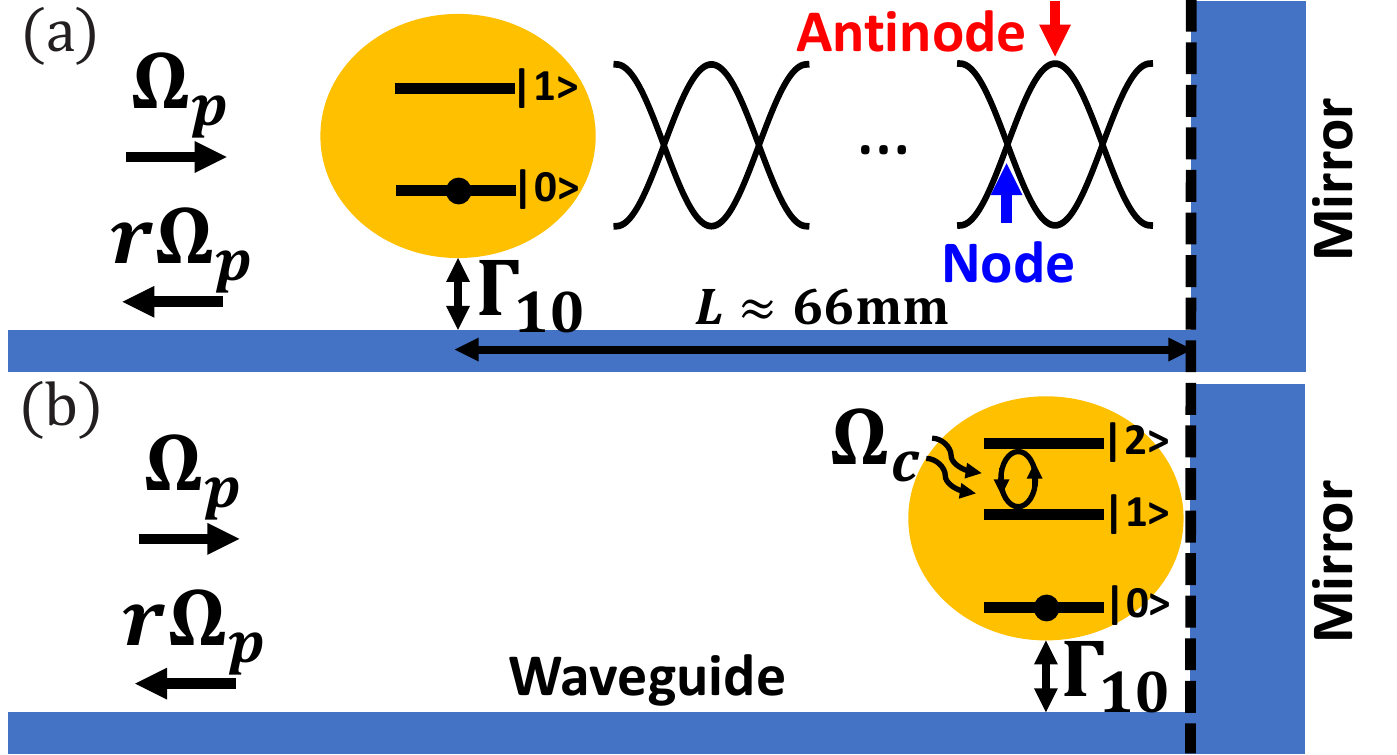}
\caption{
Illustrations of the atom--mirror systems.
For detailed experimental setups and device images, see Sec.~S1 of Ref.~\cite{SupMat}.
The probe tone $\Omega_p$ can be either continuous-wave (CW) or pulsed.
The reflected signal $r \Omega_p$ is measured by either a vector network analyzer or a digitizer.
(a) Setup for Device 1, where $\Gamma_{10}$ is tuned by changing the atom frequency.
(b) Setup for Device 2, where $\Gamma^n_{10}$ is tuned by CW pumping with $\Omega_c$ on the $\ket{1} \leftrightarrow \ket{2}$ transition.
}
\label{fig:setup}
\end{figure}

We demonstrate this switching between positive and negative $\tau_d$ using two different methods (controlling either $\Gamma_{10}$ or $\Gamma_{10}^n$) in two separate devices, as illustrated in \figref{fig:setup}.
The first device [Device 1, \figpanel{fig:setup}{a}] has a two-level atom located a distance $L \approx \unit[66]{mm}$ away from the end of the TL, which acts as an effective mirror~\cite{lu2021propagating, PRXQuantum.3.020305, lu2021quantum}.
We chose a sufficiently short $L$ to keep the system Markovian ($2 L / v_g \ll \Gamma_{10}^{-1}$).
The incident light mode and the mode reflected from the mirror can therefore be considered to interact almost simultaneously with the atom.
The mirror creates a standing-wave pattern with nodes and antinodes along the TL.
With the atom at an antinode, the atom--photon coupling, and thus $\Gamma_{10}$, is maximized, causing a resonant weak incident field to be reflected with a phase shift of $\pi$, akin to ground reflection.
Conversely, when the atom is placed at a node, it decouples from the TL and becomes transparent to the resonant TL mode, i.e., $\Gamma_{10} = 0$.
By changing a magnetic flux, we can tune the $\ket{0} \leftrightarrow \ket{1}$ transition frequency $\omega_{10}$ of the atom.
In this way, we can move the atom between nodes and antinodes, thereby adjusting $\Gamma_{10}$, as has been demonstrated in Ref.~\cite{Hoi2015}.
For $\omega_{10}$ yielding $\Gamma_{10} / 2 > \Gamma_{10}^n$, the system exhibits positive $\tau_d$; when $\Gamma_{10} / 2 < \Gamma_{10}^n$, we observe negative $\tau_d$.

The second device [Device 2, \figpanel{fig:setup}{b}] is a three-level atom positioned right next to the mirror ($L = 0$)~\cite{Lin2022, Cheng2023}.
With this placement, the atom is at an antinode of all TL modes, including both the probe and control tones.
Compared to Device 1, the atom--photon coupling is designed to be weaker in Device 2 (see Fig.~S1 in the Supplementary Material~\cite{SupMat}) to decrease the atom linewidth $\gamma_{10}$ ($\ket{0} \leftrightarrow \ket{1}$ decoherence rate), and $\Gamma^n_{10}$ is suppressed.
As a result, Device 2 has high coherence and a steep phase dispersion profile, enhancing the control of the light velocity compared to Device 1.
In contrast to Device 1, we here control $\Gamma_{10}^n$ rather than $\Gamma_{10}$.
We do so by utilizing Autler--Townes splitting (ATS)~\cite{Autler1955}: we increase $\Gamma_{10}^n$ by pumping at the $\ket{1} \leftrightarrow \ket{2}$ transition.


\textit{Measurement.}---We investigate $\tau_d$ in both the frequency and the time domains.
The detailed experimental setup is provided in Sec.~S1 of Ref.~\cite{SupMat}.
To observe light changing speed, we use a probe tone with Rabi frequency $\Omega_p$ significantly weaker than $\gamma_{10}$.
This ensures weak excitation, i.e., the atom mostly resides in its ground state.

In the frequency domain, we use a vector network analyzer to measure the system's $r$ near $\omega_{10}$.
This allows us to extract atomic parameters, as described in Secs.~S2 and S6 of Ref.~\cite{SupMat}.
For a weak drive with $\Omega_p \ll \gamma_{10}$,
\be
r = 1 - \frac{\Gamma_{10}}{\gamma_{10} + i \delta_{p,10}} ,
\label{eq:refl_coeff}
\ee
where $\delta_{p,10} = \omega_p - \omega_{10}$ is the detuning between the probe frequency $\omega_p$ and $\omega_{10}$.
At $\delta_{p,10} = 0$, the group delay time of the light is given by Eq.~(S29) in Ref.~\cite{SupMat}:
\be
\mleft. \tau_d \mright |_{\delta_{p,10} = 0} = - \mleft. \frac{\partial \phase{r}}{\partial \omega_p} \mright |_{\delta_{p,10} = 0} = \gamma_{10}^{-1} \frac{\Gamma_{10}}{\frac{\Gamma_{10}}{2} - \Gamma_{10}^n},
\label{eq:delay_two_tone}
\ee
where $\Gamma^n_{10} = \gamma_{10} - \Gamma_{10} / 2$ is the sum of the pure dephasing and the intrinsic loss of the $\ket{0}$ and $\ket{1}$ states~\cite{Hoi2013Microwave, Lu2021}.
The sign of $\tau_d$ in \eqref{eq:delay_two_tone} depends on whether $\Gamma_{10}^n$ or $\Gamma_{10} / 2$ dominates in the denominator.
Therefore, this competition between decoherence mechanisms determines whether positive or negative $\tau_d$ is observed.

We verify $\tau_d$ in the time domain by measuring the time difference between cases with and without atom interaction using a probing Gaussian pulse, $\Omega_p \exp \mleft[ - t^2 / \mleft( 2 \sigma^2 \mright) \mright]\cos{\omega_{p} t}$, generated by an amplitude-modulated radio-frequency source.
The choice of the pulse width $\sigma > \gamma_{10}^{-1}$ ensures that the pulse falls within a range $\gamma_{10}$ of $\omega_{10}$ in the frequency domain.
As a result, the output is the delayed/advanced version of the input and is also rescaled by $r$, as discussed in Sec.~S7 of Ref.~\cite{SupMat}.
The effect of different $\sigma$ with respect to $\gamma_{10}$ is discussed in Sec.~S3 of Ref.~\cite{SupMat}.
To establish the reference case without interaction, we detune the atom far away from the probe tone using an external magnetic flux.


\begin{figure*}
\includegraphics[width=\linewidth]{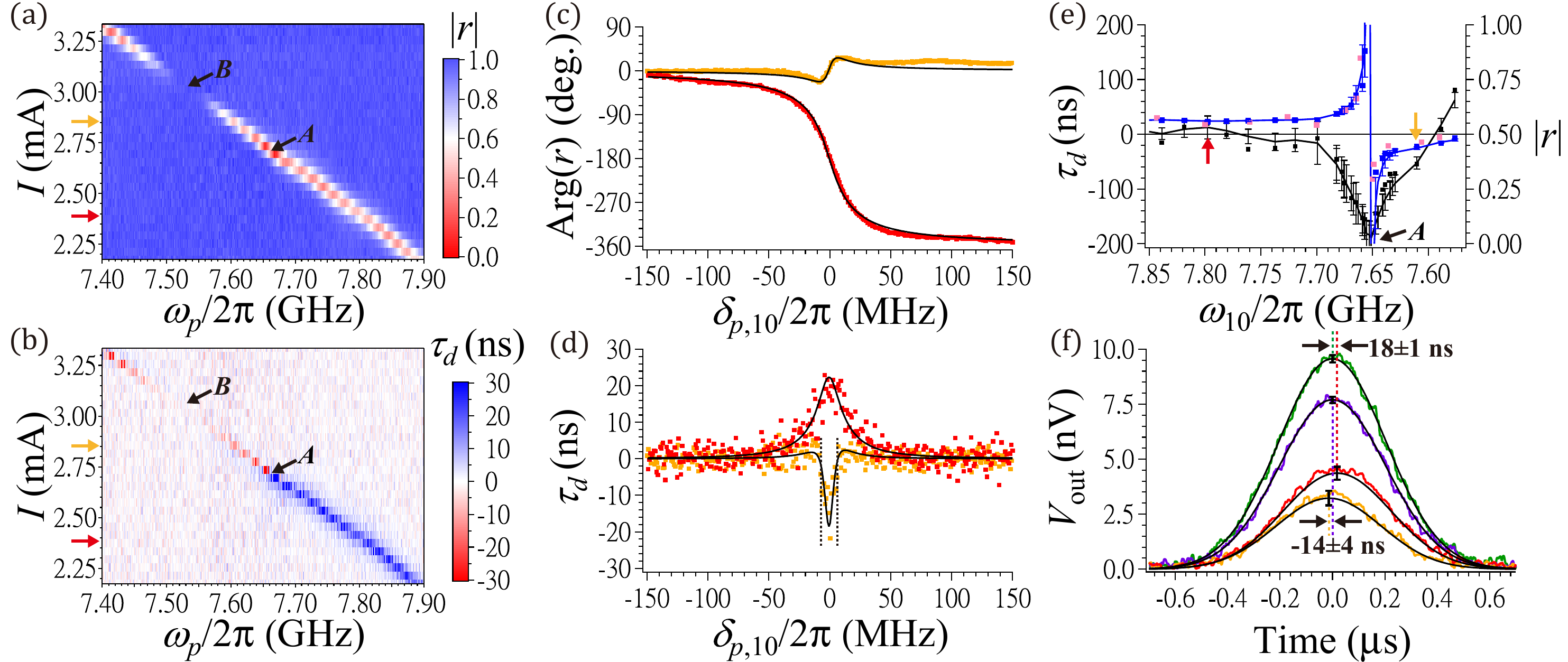}
\caption{
Tuning $\Gamma_{10}$ in Device 1 to modify $\tau_d$. 
Red arrows and data points indicate an example of positive $\tau_d$ at $\omega_{10} / 2\pi = \unit[7.7990]{GHz}$, while the orange arrows and data points show an example of negative $\tau_d$ at $\omega_{10} / 2\pi = \unit[7.6057]{GHz}$. 
Solid curves represent theoretical simulations based on \eqsref{eq:refl_coeff}{eq:delay_two_tone} in panels (c)--(e) and the optical Bloch equation (provided in Ref.~\cite{Lin2022}) in panel (f), using extracted parameters given in Tables S1 and S2 of Ref.~\cite{SupMat}.
Arrow $A$ indicates the singularity of $\tau_d$ at $\unit[7.650]{GHz}$, while arrow $B$ corresponds to the node at $\unit[7.534]{GHz}$, where $|r| = 1$ and $\tau_d$ vanishes.
(a) Measured $|r|$ as a function of bias current $I$ (magnetic field) and probe frequency $\omega_p$, where $\omega_{10}$ is set by $I$.
(b) Numerically calculated $\tau_d$ from the measured $\phase{r}$ using \eqref{eq:delay_two_tone}.
Colored arrows indicate line cuts corresponding to data points in (c,d).
(c,d) Line cuts of $\phase{r}$ and $\tau_d$, respectively, for different $\omega_{10}$.
We observe a sign change between the slopes of the red and orange $\phase{r}$ curves at $\delta_{p,10} = 0$, indicating the switching between positive and negative $\tau_d$.
(e) $|r|$ (black data points) and $\tau_d$ (blue data points) versus $\omega_{10}$ when $\delta_{p,10} = 0$.
These values are obtained from fine-grained measurements conducted near the singularity [arrow A in (a,b)].
The extracted $\tau_d$ from time-domain measurements are shown as pink data points.
(f) Time-domain results.
The green (purple) curve serves as the reference for the red (orange) curve, measured under far-detuned conditions.
}
\label{fig:natural_linewidth_result}
\end{figure*}

\textit{Results.}---We start by examining the impact on $\tau_d$ of tuning $\Gamma_{10}$. 
In Device 1, using the setup illustrated in \figpanel{fig:setup}{a} and adjusting $\omega_{10}$ with an external magnetic flux controlled by a current $I$ [see \figpanel{fig:natural_linewidth_result}{a}], we manipulate $\Gamma_{10}$ to modify $\tau_d$ according to \eqref{eq:delay_two_tone}.
The dependencies of $\Gamma_{10}$ and $\Gamma_{10}^n$ on $\omega_{10}$ can be found in Fig.~S3(c) of Ref.~\cite{SupMat}; they are extracted from \figpanel{fig:natural_linewidth_result}{a} and the corresponding phase response using \eqref{eq:refl_coeff}. 

In \figpanel{fig:natural_linewidth_result}{b}, $\tau_d$ [the derivative of $\phase{r}$; see \eqref{eq:delay_two_tone}] exhibits two distinct regions, separated by a singularity indicated by arrow $A$ at $\unit[7.650]{GHz}$, where $\Gamma_{10} / 2 = \Gamma_{10}^n$.
Between $\unit[7.650]{GHz}$ and $\unit[7.900]{GHz}$, where $\Gamma_{10} / 2 > \Gamma_{10}^n$, moving away from the node of the electric field indicated by arrow B enhances $\Gamma_{10}$~\cite{Hoi2015}, resulting in the emergence of a blue region with positive $\tau_d$ in \figpanel{fig:natural_linewidth_result}{b}.
The typical $\phase{r}$ and $\tau_d$ line cuts indicated by the red arrow in \figpanel{fig:natural_linewidth_result}{b}, showing a normal dispersion, are depicted by red data points in \figpanel{fig:natural_linewidth_result}{c,d}, respectively.

At the singularity, where $\Gamma_{10} / 2 = \Gamma_{10}^n$ and $\delta_{p,10} = 0$, the coherently reflected energy vanishes ($r = 0$) as shown in \figpanel{fig:natural_linewidth_result}{a,e} (arrow $A$) and in accordance with \eqref{eq:refl_coeff}.
All the input photons are either scattered incoherently due to the pure dephasing or converted into intrinsic loss, rendering $\tau_d$ undefined. 
This behavior is most clearly seen at arrow $A$ in \figpanel{fig:natural_linewidth_result}{e}.
Notably, $|\tau_d|$ experiences a significant increase near the singularity compared to other regions.
This increase arises from the competition between $\Gamma_{10} / 2$ and $\Gamma_{10}^n$ in the denominator of \eqref{eq:delay_two_tone}.

As we move $\omega_{10}$ from the singularity towards the node at $\unit[7.534]{GHz}$ [arrow $B$ in \figpanel{fig:natural_linewidth_result}{a,b}], $\Gamma_{10}$ is weakened due to approaching the node of the electric field~\cite{Hoi2015}.
This results in $\Gamma_{10} / 2 < \Gamma_{10}^n$ and the emergence of a red region with negative $\tau_d$ near $\delta_{p,10} = 0$ in \figpanel{fig:natural_linewidth_result}{b}.
The orange data points in \figpanel{fig:natural_linewidth_result}{c,d} represent typical line cuts of $\phase{r}$ and $\tau_d$ in this region, extracted at the orange arrow in \figpanel{fig:natural_linewidth_result}{b}, demonstrating anomalous dispersion.
In \figpanel{fig:natural_linewidth_result}{d}, we observe that this region can be further divided along the $\delta_{p,10}$ axis, with the boundary (dotted lines) defined by $\tau_d = 0$ separating the off-resonance positive $\tau_d$ from the near-resonance negative $\tau_d$ (with a width of $2 \sqrt{\gamma_{10} \mleft( \Gamma_{10}^n - \Gamma_{10} / 2 \mright)}$).
In the positive $\tau_d$ region, the dominance of $\Gamma_{10}^n$ is attenuated by $\delta_{p,10}$, as described by the general $\tau_d$ expression in Eq.~(S29) in Sec.~S8 of Ref.~\cite{SupMat}.
This results in $\Gamma_{10} / 2$ domination and consequently a weakly positive $\tau_d$.
As $\delta_{p,10}$ approaches zero, the probe tone becomes more susceptible to $\Gamma_{10}^n$, turning from positive to negative $\tau_d$.
As $\omega_{10}$ approaches the node, $|\tau_d|$ diminishes due to weak interaction with the TL.
This is indicated by the numerator of \eqref{eq:delay_two_tone} and shown on the right side of the arrow $A$ in \figpanel{fig:natural_linewidth_result}{e}.

In \figpanel{fig:natural_linewidth_result}{f}, we verify the time-domain $\tau_d$ at different $\omega_{10}$.
We observe $\tau_d = \unit[18]{ns}$ at $\unit[7.799]{GHz}$ (red) and $\tau_d = \unit[-14]{ns}$ at $\unit[7.6057]{GHz}$ (orange).
The complete extracted time-domain $\tau_d$ are shown as pink data points in \figpanel{fig:natural_linewidth_result}{e}; they match the frequency-domain results (blue data points).
In \figpanel{fig:natural_linewidth_result}{e}, the roughness in $|r|$ can be attributed to uncontrolled background resonances that spread across the spectrum.
These resonances weakly interfere with the atom spectrum and lead to errors in the extracted parameters (see Fig.~S3(c) of Ref.~\cite{SupMat}).
As a result, the simulated $|r|$ [solid black curve, calculated using \eqref{eq:refl_coeff} and the extracted parameters in Fig.~S3(c) of Ref.~\cite{SupMat}] deviates from the measured black data points.

\begin{figure*}
\includegraphics[width=\linewidth]{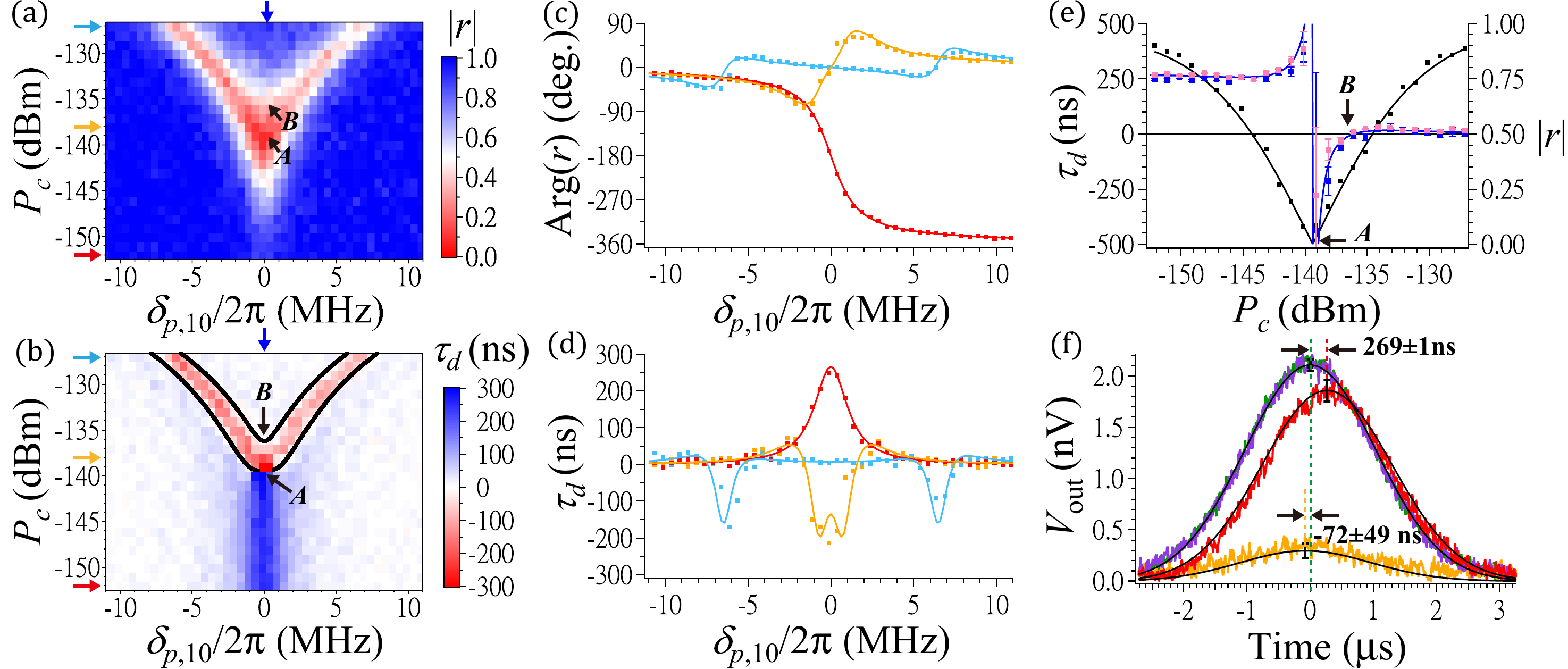}
\caption{
Tuning $\tau_d$ through pump-induced $\Gamma_{10}^n$ in Device 2.
Deep blue arrows indicates the $\delta_{p,10} = 0$ vertical line cut.
The arrows $A$ and $B$ indicate the location of the singularity ($r = 0$) at $\unit[-139.4]{dBm}$ and $\tau_d = 0$ at $\unit[-136.0]{dBm}$, respectively.
Solid curves correspond to simulations based on the two-tone reflection formula (Eq.~(S5) in Ref.~\cite{SupMat}) in (c)--(e), and formulas for time evolution (Eqs.~(S6)--(S8) and (S16) in Ref.~\cite{SupMat}) in (f).
These simulations utilize the extracted parameters given in Tables S1 and S2 in Ref.~\cite{SupMat}. 
(a) Measured $|r|$ as a function of $P_c \propto \Omega_c^2$ and $\delta_{p,10}$.
(b) Numerically calculated $\tau_d$ from the measured $\phase{r}$ according to \eqref{eq:delay_two_tone}.
The black boundary curves represent $\tau_d = 0$.
Colored horizontal arrows indicate different line cuts [$P_c = \unit[-127.1]{dBm}$ (cyan), $\unit[-138.1]{dBm}$ (orange), and $\unit[-152.1]{dBm}$ (red)] corresponding to those in (c,d); $\delta_{p,10} = 0$ (deep blue) corresponds to the one in (e).
The simulations for (a,b) are shown in Fig.~S7(b,c) of Ref.~\cite{SupMat}.
(c,d) Line cuts of $\phase{r}$ and $\tau_d$, respectively, for different $P_c$.
(e) Power dependence of $|r|$ (black data points) and $\tau_d$ (blue data points) when $\delta_{p,10} = 0$.
The extracted $\tau_d$ from time-domain measurements are shown as pink data points~\cite{note1}.
(f) Time-domain results for $P_c=\unit[-115.1]{dBm}$ (purple), $\unit[-138.1]{dBm}$ (orange), and $\unit[-152.1]{dBm}$ (red).
The green curve serves as the reference for the red and orange curves, measured under far-detuned condition.
Additional details for different $P_c$ are provided in Fig.~S7(d,e) of Ref.~\cite{SupMat}.
}
\label{fig:photon_assisted_linewidth_result}
\end{figure*}

Next, we focus on pump-induced $\Gamma _{10}^n$.
In Device 2, a control tone (with Rabi frequency $\Omega_c$) is introduced to modify $\Gamma_{10}^n$ as illustrated in \figpanel{fig:setup}{b}.
Near resonance, $|\delta_{p,10}| < \gamma_{10}$, the three-level atom can be treated as an effective two-level system by replacing $\Gamma_{10}$, $\gamma_{10}$, and $\Gamma^n_{10}$ in \eqsref{eq:refl_coeff}{eq:delay_two_tone} with the following effective rates (see Eq.~(S24) in Ref.~\cite{SupMat}), respectively:
\begin{flalign}
\Gamma &= \frac{\Gamma_{10}}{1 - \mleft( \frac{\Omega_c}{2 \gamma_{20}} \mright)^2} ,
\label{eq:eff_relax}
\\
\gamma &= \gamma_{10} \frac{1 + \frac{\Omega_c^2}{4 \gamma_{10} \gamma_{20}}}{1 - \mleft( \frac{\Omega_c}{2 \gamma_{20}} \mright)^2} ,
\label{eq:eff_decoh}
\\
\Gamma^n &= \gamma - \frac{\Gamma}{2} = \Gamma^n_{10} \frac{1 + \frac{\Omega_c^2}{4 \Gamma^n_{10} \gamma_{20}}}{1 - \mleft( \frac{\Omega_c}{2 \gamma_{20}} \mright)^2} .
\label{eq:eff_dephase}
\end{flalign}
Here, $\gamma_{20}$ is the $\ket{0} \leftrightarrow \ket{2}$ decoherence rate.
The numerator in \eqsref{eq:eff_decoh}{eq:eff_dephase} captures the power broadening.
In \eqref{eq:eff_dephase}, the presence of the control tone adds additional non-radiative decay on top of $\Gamma^n_{10}$, allowing us to adjust $\Gamma^n$.
In the absence of $\Omega_c$, the narrow linewidth of the system results in $\tau_d = \unit[275]{ns}$ for a resonant probe.
The detailed characterization of Device 2 is presented in Fig.~S2 with extracted parameters in Table S1~\cite{SupMat}.

In \figpanel{fig:photon_assisted_linewidth_result}{a,b}, we sweep the control-tone power $P_c$.
At $P_c = \unit[-139.4]{dBm}$ ($\Omega_c \approx 2\pi \cdot \unit[3.276]{MHz}$) in \figpanel{fig:photon_assisted_linewidth_result}{b}, we observe a transition from positive $\tau_d$ (blue data points) to negative $\tau_d$ (red data points) at the singularity indicated by arrow $A$.
The singularity corresponds to $\Gamma / 2 = \Gamma^n$, where $\Omega_c = 2 \sqrt{\gamma_{20} \mleft( \Gamma_{10} - \gamma _{10} \mright)}$ according to \eqsrefCross{eq:eff_relax}{eq:eff_dephase}.
The transition can also be observed from the vertical line cut in \figpanel{fig:photon_assisted_linewidth_result}{a,b}, indicated by the deep blue arrows at $\delta_{p,10} = 0$, shown in \figpanel{fig:photon_assisted_linewidth_result}{e}.
Typical normal (anomalous) dispersion of $\phase{r}$ and $\tau_d$ is illustrated by the red (orange) data points in \figpanel{fig:photon_assisted_linewidth_result}{c,d}, respectively, exhibiting positive (negative) $\tau_d$ at $\delta_{p,10} = 0$.

Note the similarity between the region $\Omega_c < 2 \gamma_{20}$ in \figpanel{fig:photon_assisted_linewidth_result}{c,d,e} and the results in \figpanel{fig:natural_linewidth_result}{c,d,e}.
In \figpanel{fig:photon_assisted_linewidth_result}{d}, the presence of two dips in the orange trace near $\delta_{p,10} = 0$ indicates the ATS.
The behavior of $\tau_d$ starts to deviate from the case of tuning $\Gamma_{10}$ when it approaches $P_c = \unit[-136.0]{dBm}$, where $\Omega_c = 2 \gamma_{20}$ [marked by arrow $B$ in \figpanel{fig:photon_assisted_linewidth_result}{a,b,e}].
In \figpanel{fig:photon_assisted_linewidth_result}{a,b}, the atom exhibits a splitting into two transitions due to ATS.
As depicted by the cyan trace in \figpanel{fig:photon_assisted_linewidth_result}{c,d}, each transition displays a relatively weaker anomalous dispersion profile.
In \figpanel{fig:photon_assisted_linewidth_result}{f}, we showcase the ability to switch between different regions along $\delta_{p,10} = 0$.
The $P_c = \unit[-152.1]{dBm}$ (red) trace corresponds to $\tau_d = \unit[269]{ns}$, whereas the $P_c = \unit[-138.1]{dBm}$ (orange) trace represents $\tau_d = \unit[-72]{ns}$.
At $P_c = \unit[-115.1]{dBm}$ (purple trace), the ATS becomes wide enough for the atom to be far-off resonance with the probe tone.
Consequently, the group delay is completely suppressed and effectively switched off.


\textit{Conclusion.}---By controlling radiative decay or pump-induced non-radiative decay in two devices with an artificial atom in front of a mirror, we showed that when half of the radiative decay rate is greater (smaller) than non-radiative decay rate, we observe positive (negative) group delay for light interacting with the artificial atom.
Both switching methods enable an on-demand and in-situ transition between positive and negative group delay, giving control over the group velocity of light.
Our results are an important step for signal manipulation at the single-photon level in waveguide QED.


\begin{acknowledgments}

I.-C.H.~acknowledges financial support from City University of Hong Kong through the start-up project 9610569 and from the Research Grants Council of Hong Kong (Grant No.~11312322).
G.D.L.~and K.T.L.~thank MOST/NSTC of Taiwan under Grant Nos.~112-2112-M-002-001 and 112-2811-M-002-067.
A.F.K.~acknowledges support from the Swedish Research Council (grant number 2019-03696), the Swedish Foundation for Strategic Research (grant numbers FFL21-0279 and FUS21-0063), the Horizon Europe programme HORIZON-CL4-2022-QUANTUM-01-SGA via the project 101113946 OpenSuperQPlus100, and from the Knut and Alice Wallenberg Foundation through the Wallenberg Centre for Quantum Technology (WACQT).
Z.-R.L.~acknowledges support from Shanghai Technology Innovation Action Plan Integrated Circuit Technology Support Program (No.~22DZ1100200).

\end{acknowledgments}


\bibliography{main}

\begin{thebibliography}{50}
\bibitem [{not()}]{note1}%
  \BibitemOpen
  \href@noop {} {}\bibinfo {note} {Near the singularity at point $A$, where the slope of $\tau_d$ is large, errors in $P_c$ and atomic parameters can amplify the error in $\tau_d$. Another reason is that the bandwidth of $\tau_d$ narrows when the operating point approaches the singularity. This width can be potentially smaller than $\sigma^{-1}$ even if $\sigma^{-1} \ll \gamma_{10}$ is met. These factors contribute to the disagreement between the time- and frequency-domain results near the singularity.}\BibitemShut {Stop}%
\end{thebibliography}
\end{document}


\title{Supplementary Material for \\``Group delay controlled by the decoherence of a single artificial atom"}

\author{Y.-T.~Cheng}
\thanks{These authors contributed equally}
\affiliation{Department of Physics, City University of Hong Kong, Kowloon, Hong Kong SAR 999077, China}

\author{K.-M.~Hsieh}
\thanks{These authors contributed equally}
\affiliation{Department of Physics, City University of Hong Kong, Kowloon, Hong Kong SAR 999077, China}

\author{B.-Y.~Wu}
\thanks{These authors contributed equally}
\affiliation{Department of Physics, City University of Hong Kong, Kowloon, Hong Kong SAR 999077, China}

\author{Z.~Q.~Niu}
\thanks{These authors contributed equally}
\affiliation{State Key Laboratory of Materials for Integrated Circuits, Shanghai Institute of Microsystem and Information Technology (SIMIT), Chinese Academy of Sciences, Shanghai 200050, China}
\affiliation{ShanghaiTech University, Shanghai 201210, China}

\author{F.~Aziz}
\affiliation{Department of Physics, National Tsing Hua University, Hsinchu 30013, Taiwan}

\author{Y.-H.~Huang}
\affiliation{Department of Physics, National Tsing Hua University, Hsinchu 30013, Taiwan}

\author{P.~Y.~Wen}
\affiliation{Department of Physics, National Chung Cheng University, Chiayi 621301, Taiwan}

\author{K.-T.~Lin}
\affiliation{CQSE, Department of Physics, National Taiwan University, Taipei 10617, Taiwan}

\author{Y.-H. Lin}
\affiliation{Department of Physics, National Tsing Hua University, Hsinchu 30013, Taiwan}
\affiliation{Center for Quantum Technology, National Tsing Hua University, Hsinchu 30013, Taiwan}

\author{J.~C.~Chen}
\affiliation{Department of Physics, National Tsing Hua University, Hsinchu 30013, Taiwan}
\affiliation{Center for Quantum Technology, National Tsing Hua University, Hsinchu 30013, Taiwan}

\author{A.~F.~Kockum}
\affiliation{Department of Microtechnology and Nanoscience (MC2), Chalmers University of Technology, SE-412 96 Gothenburg, Sweden}

\author{G.-D.~Lin}
\affiliation{CQSE, Department of Physics, National Taiwan University, Taipei 10617, Taiwan}
\affiliation{Physics Division, National Center for Theoretical Sciences, Taipei 10617, Taiwan}

\author{Z.-R.~Lin}
\email[e-mail:]{zrlin@mail.sim.ac.cn}
\affiliation{State Key Laboratory of Materials for Integrated Circuits, Shanghai Institute of Microsystem and Information Technology (SIMIT), Chinese Academy of Sciences, Shanghai 200050, China}

\author{Y.~Lu}
\email[e-mail:]{kdluyong@outlook.com}
\affiliation{Guangzhou Institute of technology, Xidian University, Xi'an, China}
\affiliation{Advanced Interdisciplinary Research Center, Xidian University, Xi'an, China}
\affiliation{Faculty of Integrated Circuit, Xidian University, Xi'an, China}

\author{I.-C.~Hoi}
\email[e-mail:]{iochoi@cityu.edu.hk}
\affiliation{Department of Physics, City University of Hong Kong, Kowloon, Hong Kong SAR 999077, China}

\date{\today}

\maketitle

\tableofcontents


\section{Experimental setup}

\begin{figure}
\includegraphics[width=\linewidth]{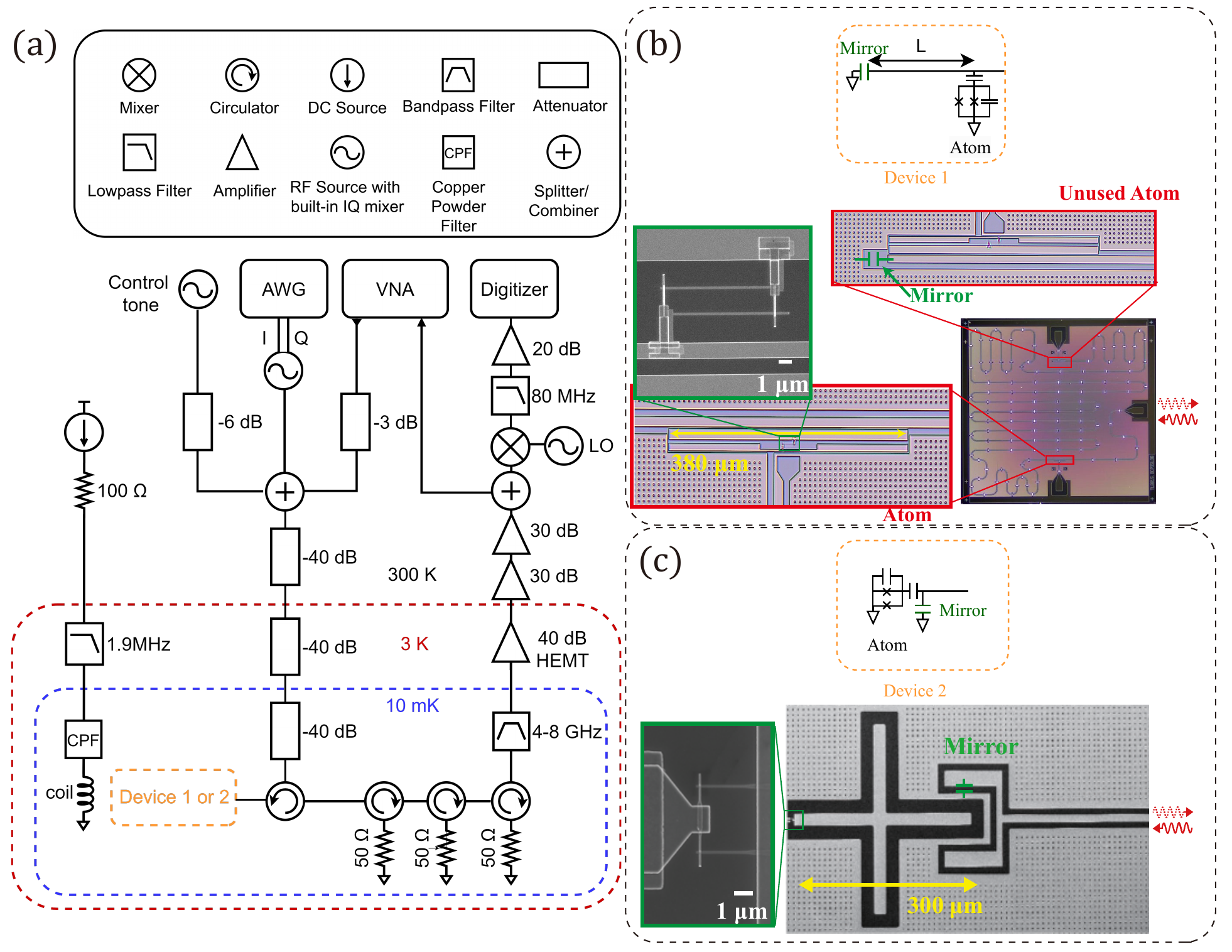}
\caption{
Measurement setup and device images.
(a) The frequency- and time-domain measurement systems are combined together using splitters.
The device under test (orange dashed box) is positioned at the bottom of the dilution refrigerator, operating at a base temperature of \unit[10]{mK} (blue dashed box).
(b) Optical microscope and scanning electron microscope (SEM) images of Device 1:
the transmon-type artificial atoms~\cite{Koch2007} are highlighted by the red boxes with enlarged views, while an SEM image displays the Josephson junctions forming a superconducting quantum interference device (SQUID), depicted in the green box.
The capacitance at the end of the transmission line serves as a mirror.
The orange dashed box contains the equivalent circuit of the device.
(c) Optical microscope and SEM images of Device 2:
the transmon-type artificial atom is capacitively coupled to the end of the transmission line as shown in the image.
An SEM image displays the Josephson junctions forming a SQUID, indicated in the green box.
The atom is positioned at the mirror with $L = 0$, ensuring that the atom is at the antinodes of electric fields in the waveguide.
}
\label{fig:exp_setup}
\end{figure}

The experimental setups used for conducting the measurements described in the main text are shown in \figref{fig:exp_setup}.
For the frequency-domain measurement, a vector network analyzer (VNA) is employed to transmit and receive the continuous-wave (CW) signal.
The time-domain measurement instead employs an in-phase-and-quadrature (IQ) modulated radiofrequency (RF) source driven by an arbitrary-waveform generator (AWG) serving as the transmitter, while a digitizer, along with a down-converting mixer and an RF source, functions as a heterodyne receiver.
To combine the frequency- and time-domain measurement systems, two splitters are used at the input and output ports of the dilution refrigerator (DR) at room temperature.
These systems generate the probe tone for different types of measurements.
Furthermore, an RF source acts as a control tone and is combined with the probe tone using the same splitter functioning as a combiner at the input port of the DR.

The combined input signal is directed towards the device, which consists of the atom-mirror system (orange dashed box) situated in a dilution refrigerator (blue dashed box) at a base temperature of $T = \unit[10]{mK}$.
This low temperature ensures that the atom has a negligible thermal population ($k_B T \ll \hbar \omega_{10}$, where $k_B$ is Boltzmann's constant) and therefore is in its ground state when an experiment begins.
After interacting with the atom, the reflected signal undergoes amplification and filtering processes before returning to room temperature.
The signal is split in half by the splitter and directed to the receiver parts of the two measurement systems separately.

In the DR, two devices are measured individually and undergo identical characterization procedures.
Device 1 [see \figpanel{fig:exp_setup}{b}] incorporates an artificial atom positioned at a distance $L \approx \unit[66]{mm}$ from the mirror.
By adjusting the global magnetic flux to set the flux through the superconducting quantum interference device (SQUID) formed by the Josephson junctions of the artificial atom, we modify the transition frequencies $\omega_{10}$ such that the atom can be selectively positioned at either an antinode or a node of the resonant electric field.
In contrast, Device 2 [see \figpanel{fig:exp_setup}{c}] features an artificial atom positioned at the mirror ($L = 0$), which always is at the antinode of the electric field.


\section{Reflection coefficient as a function of probe power and probe frequency}

In this section, we first perform reflection spectroscopy with a single tone on Devices 1 and 2 to extract parameters of the $\ket{0} \leftrightarrow \ket{1}$ transition: the transition frequency $\omega_{10}$, the relaxation rate $\Gamma_{10}$, the decoherence rate $\gamma_{10}$, and the atom-field coupling constant $k_{10}$. 
Next, we investigate the phenomenon of positive and negative group delays using a weak continuous probe in the frequency domain.

For a continuous probe of frequency $\omega_p$ interacting with an atom in front of a mirror, the reflection coefficient is given by~\cite{hoi2013thesis}
\be
r = 1 - \frac{\Gamma_{10}}{\gamma_{10}} \frac{1 - i \mleft( \delta_{p,10} / \gamma_{10} \mright)}{1 + \delta_{p,10}^2 / \gamma_{10}^2 + \Omega_p^2 / \Gamma_{10} \gamma_{10}} ,
\label{eq:r}
\ee
where $\delta_{p,10} = \omega_p - \omega_{10}$ is the detuning between the probe frequency and the atom resonance frequency, and $\Omega_p$ is the Rabi frequency of the probe, which is proportional to the voltage amplitude $V_p$ of the probe, or, equivalently, to the square root of chip-level power $P_p$, i.e., $\Omega_p = k_{10} \sqrt{P_p}$.
The non-radiative decay rate of the atom is
\be
\Gamma^n_{10} = \gamma_{10} - \Gamma_{10} / 2,
\label{eq:puredephasing}
\ee
and is defined to contain both pure dephasing and intrinsic loss (i.e., decay to other environments than the transmission line).

If a weak probe ($\Omega_p \ll \gamma_{10}$) is used, \eqref{eq:r} becomes Eq.~(1) in the main text.
Furthermore, if the probe also is resonant ($\delta_{p,10} = 0$), \eqref{eq:r} becomes
\be
r = 1 - \frac{\Gamma_{10}}{\gamma_{10}}.
\label{eq:strongcoupling}
\ee
In the case of strong coupling, where $\Gamma_{10} \gg \Gamma^n_{10}$, \eqref{eq:puredephasing} is simplified and incorporated into \eqref{eq:strongcoupling}, resulting in $r = -1$.
It thus turns out that a resonant weak probe field is fully reflected with a $\pi$ phase shift.

\begin{figure}
\includegraphics[width=\linewidth]{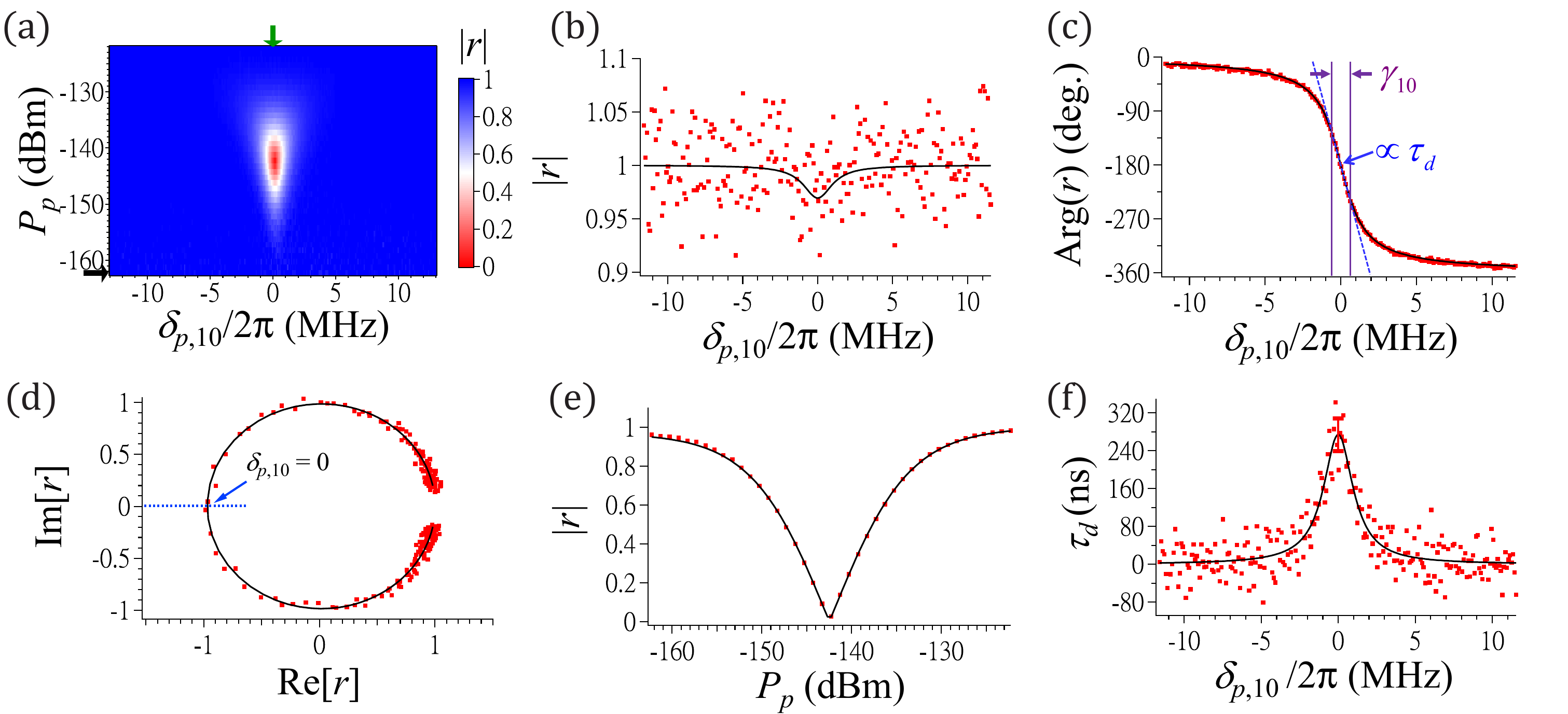}
\caption{
Characterization of the $\ket{0} \leftrightarrow \ket{1}$ transition of Device 2.
(a) Reflection magnitude $|r|$ response as a function of detuning frequency ($\delta_{p,10} / 2\pi$) and probe power $P_p$.
(b) Horizontal line cut at the position marked by the black arrow in (a).
(c) Phase response of the line cut in (b). 
(d) Reflection coefficient plotted in the IQ plane for a weak probe ($\Omega_p / 2\pi = \unit[0.166]{MHz}$).
(e) Vertical line cut at the position marked by the green arrow in (a). 
(f) Calculated group delay time from (c) using \eqref{eq:delay}.
On resonance, a $\pi$ phase shift within a small linewidth $\gamma_{10}$ results in a steep slope and therefore a high $\tau_d$.
The red dots are the measured data points.
The black curves in the figures represent theoretical fits obtained using \eqref{eq:r} [for curves in (b-e)] and \eqref{eq:delay} [for curve in (f)] with extracted parameters listed in \tabref{tab:qubit_param}.
\label{fig:s2_power_dep}
}
\end{figure}

During the calibration process for Device 2, we select the lowest power from \figpanel{fig:s2_power_dep}{a} (black arrow); this line cut is shown in \figpanel{fig:s2_power_dep}{b}.
The incident field is here almost fully reflected by the atom, with $|r| \simeq 0.97$, which indicates that $\Omega_p \ll \gamma_{10}$ and $\Gamma_{10} \gg \Gamma^n_{10}$.
Due to the low signal-to-noise ratio of the weak probe, there is a significant fluctuation in the magnitude of $|r|$.
However, when mapping $r$ in the IQ plane, as shown in \figpanel{fig:s2_power_dep}{d}, this fluctuation has a relatively lesser impact.
This enables us to employ a circle-fit method~\cite{Lu2021, Probst2015}, which is less affected by the fluctuations and provides a more robust estimation of the parameters.
By fitting the data points on the IQ plane to a circle, we can accurately determine the relevant parameters despite the presence of noise in $|r|$.
Utilizing this method, we extract the parameters $\Gamma_{10}$ and $\gamma_{10}$ using \eqref{eq:r}; the results are summarized in \tabref{tab:qubit_param} together with other extracted parameters.

\begin{table*}
\begin{tabular}{| c | c | c | c | c | c | c | c | c | }
    \hline
    Device&
    $\omega_{10} / 2 \pi$&
    $\Gamma_{10} / 2 \pi$&
    $\gamma_{10} / 2 \pi$&
    $\Gamma_{10}^n / 2 \pi$&
    $k_{10}$&
    $\Gamma_{21} / 2 \pi$&
    $\gamma_{20} / 2 \pi$&
    $\Gamma_{20}^n / 2 \pi$\\
    \hline
    -&
    MHz&
    MHz&
    MHz&
    MHz&
    Hz/$\sqrt{\text{W}}$&
    MHz&
    MHz&
    MHz\\
    \hline
    1a&
    7605.7 $\pm$ 0.7&
    6.96 $\pm$ 0.29&
    11.8 $\pm$ 0.8&
    8.3 $\pm$ 0.8&
    9.37 $\cdot$ $10^{14}$&
    -&
    -&
    -\\
    \hline
    1b&
    7799.0 $\pm$ 0.4&
    33.07 $\pm$ 0.28&
    22.6 $\pm$ 0.4&
    6.1 $\pm$ 0.4&
    2.0185 $\cdot$ $10^{15}$&
    -&
    -&
    -\\
    \hline
    2&
    4761.62 $\pm$ 0.013&
    2.316 $\pm$ 0.018&
    1.176 $\pm$ 0.013&
    0.017 $\pm$ 0.016&
    6.8363 $\cdot$ $10^{14}$&
    4.632 $\pm$ 0.037&
    2.364 $\pm$ 0.042&
    0.048 $\pm$ 0.046\\
    \hline
\end{tabular}
\caption{
Extracted parameters of the artificial atoms in Devices 1 and 2.
The Device labels 1a and 1b are used to refer to the typical parameters of Device 1 measured at two different $\ket{0} \leftrightarrow \ket{1}$ transition frequencies.
The label 1a corresponds to measurements taken near the node (the node is at $\unit[7534]{MHz}$) and 1b denotes measurements conducted near the antinode (the antinode is at $\unit[7982]{MHz}$).
The $\ket{2} \leftrightarrow \ket{0}$ decoherence rate is given by $\gamma_{20} = \Gamma_{21} / 2 + \Gamma^n_{20}$~\cite{Peropadre2013}, where $\Gamma_{21}$ and $\Gamma^n_{20}$ are the $\ket{2} \rightarrow \ket{1}$ decay rate and the combined non-radiative decay rate for the $\ket{2} \leftrightarrow \ket{0}$ transition, respectively.
The value of $\Gamma_{21}$ is determined by considering its relation to $\Gamma_{10}$ based on the coupling matrix element, which is twice the value of $\Gamma_{10}$.
}
\label{tab:qubit_param}
\end{table*}

In~\figpanel{fig:s2_power_dep}{e}, where the probe is resonant with the atom ($\delta_{p,10} = 0$), we fit the power dependence with \eqref{eq:r}, extract $k_{10}$ and determine $\Omega_p$ (see \tabref{tab:qubit_param}) with corresponding VNA power, thus obtaining the effective attenuation $A$ and gain $G$~\cite{cheng2023tuning}.
These calibration steps can be applied to both frequency-domain and time-domain measurements.
The extracted data ($A$, $G$) are shown in \tabref{tab:powerdependence}.
The parameter difference between the time- and frequency-domain setups is from the additional attenuation in the up-converting IQ modulator and the down-converting mixing in the time-domain setup.

\begin{table*}
\begin{tabular}{| c | c | c | c |}
    \hline
    $A_{\rm freq}$ & $G_{\rm freq}$ & $A_{\rm time}$ & $G_{\rm time}$\\
    \hline
    dB & dB & dB & dB \\
    \hline
    $132.3$ & $60.6$ & $143.7$ & $101.4$\\
    \hline
\end{tabular}
\caption{
Extracted effective attenuations $A$ and gains $G$.
The subscripts for $A$ and $G$ represent the time-domain and frequency-domain systems, respectively [see \figpanel{fig:exp_setup}{a}].
Details of the calibration method are shown in Ref.~\cite{cheng2023tuning}.
}
\label{tab:powerdependence}
\end{table*}

A $\pi$ phase shift occurs when the probe is resonant with the atom in \figpanel{fig:s2_power_dep}{c}. 
It suggests again that the atom-mirror system is in the strong coupling regime, where the coherent coupling $\Gamma_{10}$ is much larger that any loss or pure dephasing in the system, i.e.,~$\gamma_{10} \simeq \Gamma_{10} / 2$.
Even in the strong coupling regime, we maintain a narrow linewidth of $\gamma_{10} / 2\pi \simeq \unit[1]{MHz}$.
Consequently, there is a steep slope in the phase response, as shown in the blue curve in \figpanel{fig:s2_power_dep}{c}.
The slope is directly linked to the group delay time $\tau_d$, according to Ref.~\cite{Novikov2016}:
\be
\tau_d = -\frac{\partial \phase{r}}{\partial \omega_p} ,
\label{eq:delay}
\ee
where $\phase{r}$ is the phase of the complex reflection coefficient.
The calculated $\tau_d$ from \figpanel{fig:s2_power_dep}{c}, according to \eqref{eq:delay}, is presented in \figpanel{fig:s2_power_dep}{f} and provides a prediction of $\tau_d$ prior to the time-domain measurement. 
The predicted maximum $\tau_d$ is approximately $\unit[274]{ns}$.
Further details of the derivation of $\tau_d$ are presented in \secref{sec:GroupDelayAndReflection}.

\begin{figure}
\includegraphics[width=\linewidth]{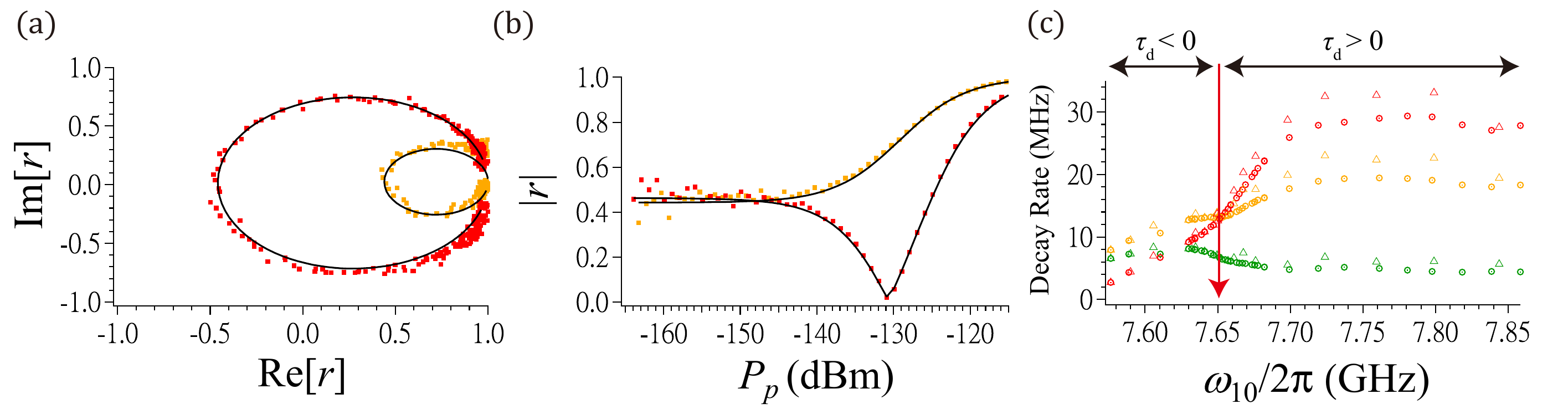}
\caption{
Characterization of Device 1: In (a) and (b), the black curves represent theoretical fittings based on \eqref{eq:r}.
The colored dots are the measured data points with the atom bias set at $\omega_{10} / 2\pi$ of $\unit[7.6057]{GHz}$ (orange) and $\unit[7.799]{GHz}$ (red), respectively.
(a) The reflection coefficient plotted in the IQ plane for a weak probe at $\Omega_p / 2\pi = \unit[0.64]{MHz}$ (orange) and $\unit[1.29]{MHz}$ (red).
(b) The on-resonance $|r|$ as a function of power for the two values of $\omega_{10} / 2\pi$. 
(c) The extracted $\Gamma_{10} / 2\pi$ (red), $\gamma_{10} / 2\pi$ (orange), and $\Gamma^n_{10} / 2\pi$ (green) when the atom is biased at different $\omega_{10} / 2\pi$.
All three rates shift slightly between two cooldowns.
The panels (a)--(b) and the spectroscopic data in Fig.~2(a)--(e) in the main text correspond to the parameters obtained from the first cooldown (represented by circles), while the time-domain-related results in Fig.~2(e)--(f) are associated with the qubit parameters obtained from the second cooldown (represented by triangles).
The red arrow indicates $\Gamma_{10} / 2 = \Gamma^n_{10}$, where a singularity occurs for the group delay time.
This separates regions of negative group delay ($\tau_d < 0$) and positive group delay region ($\tau_d > 0$).
It is important to note that the frequency of the singularity remains nearly the same between the two cooldowns.
}
\label{fig:device1}
\end{figure}

We perform the same measurements to characterize Device 1; some of the results are shown in \figref{fig:device1}.   
The diameter of the resonant circle, given by $\Gamma_{10} / \gamma_{10}$, decreases as $\omega_{10} / 2\pi$ approaches to the node frequency in \figpanel{fig:device1}{a}.
This decrease indicates that the coupling is reduced. 
We perform time-domain measurements in the subsequent sections.

\section{Group delay time as a function of pulse width}
\label{sec:delayVSsigma}

We send a weak ($\unit[-162.3]{dBm}$) probing Gaussian pulse with a variable pulse width $\sigma$ to the atom-mirror system (Device 2).
The dependence of the group delay time $\tau_d$ on the probe power is discussed in \secref{sec:delayVSprobe_power}.
The carrier frequency of the probe is provided by an RF source, and the amplitude modulation signal is produced by the AWG (see \figref{fig:exp_setup}).

\begin{figure}
\includegraphics[width=\linewidth]{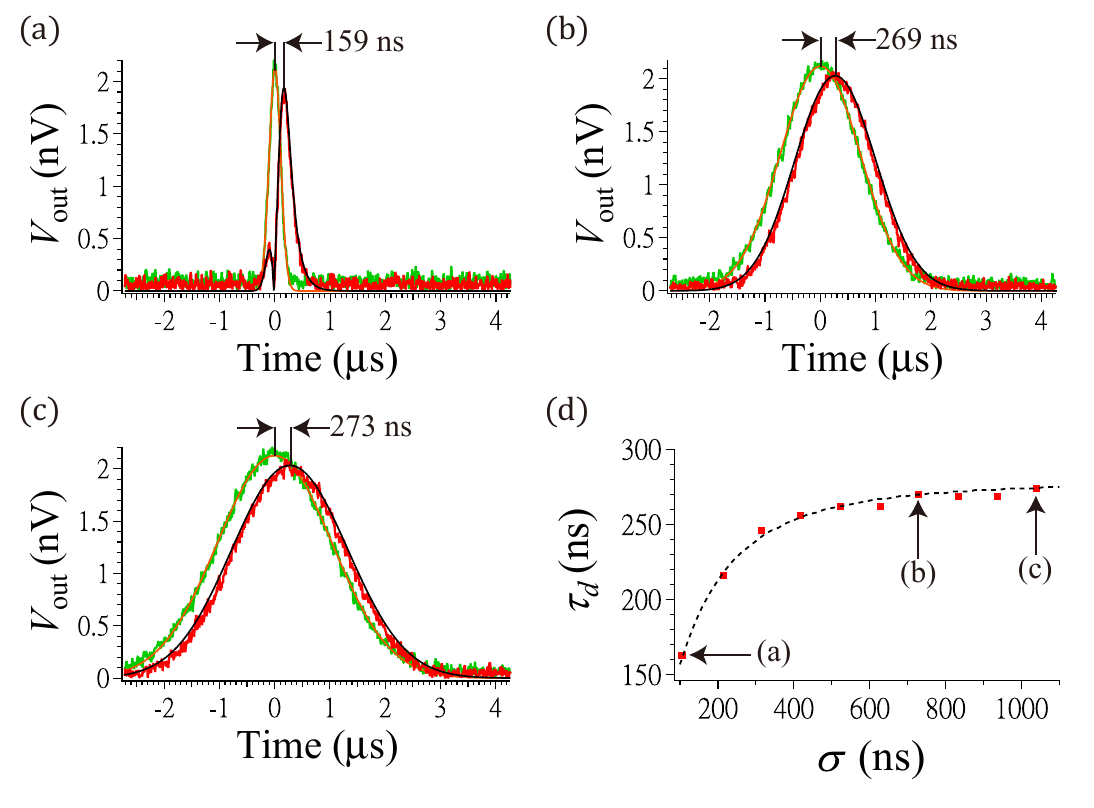}
\caption{
Sending Gaussian pulses (with a time resolution of $\unit[1]{ns}$) to the atom-mirror system for Device 2.
Pulse envelope changes of the output signals are displayed for three different pulse widths $\sigma$: (a) $\unit[105]{ns}$, (b) $\unit[728]{ns}$, and (c) $\unit[1040]{ns}$.
Experimental data are shown in red dots (with the atom) and green dots (without the atom, which then is far detuned).
The orange curves are based on Gaussian fitting, and the black curves are based on optical Bloch equations and the input-output relation~\cite{Lin2022} with the parameters of Device 2 in \tabref{tab:qubit_param}.
The theory curves maintain good agreement with the experimental data to extract $\tau_d$: (a) $\tau_d = \unit[159]{ns}$, (b) $\tau_d = \unit[269]{ns}$, and (c) $\tau_d = \unit[273]{ns}$. 
(d) Summary of $\tau_d$ as a function of $\sigma$ when the pulse is resonant with the atom.
The red dots are measured data and the dashed curve is the simulated prediction.
\label{fig:swp_sigma_delay_time}
}
\end{figure}

The theoretical simulation procedure is as follows: first, a Gaussian function is employed to fit the reference Gaussian pulse measured when the atom is far-detuned.
The parameters obtained from the fitted Gaussian pulse, represented by the orange curves in \figref{fig:swp_sigma_delay_time}, are then combined with the parameters given in \tabref{tab:qubit_param}. 
Subsequently, the evolution of the output pulses is simulated using optical Bloch equations and the input-output relation~\cite{Lin2022}.
The simulated results of the output pulse are depicted as the black solid curves in \figref{fig:swp_sigma_delay_time}.
Finally, $\tau_d$ is calculated as the time difference between the two peaks.

As depicted in Fig.~\ref{fig:swp_sigma_delay_time}, the simulation curves exhibit good agreement with the experimental data (red dots). As the value of $\sigma$ increases, $\tau_d$ increases.
The larger $\sigma$ indicates a narrower signal bandwidth distribution in frequency domain.
When $\sigma$ is $\unit[1040]{ns}$ [see \figpanel{fig:swp_sigma_delay_time}{c}], $\tau_d$ reaches up to \unit[273]{ns}.
In this scenario, the bandwidth of the incoming field ($\simeq \unit[1]{MHz}$) lies within the linewidth of the artificial atom ($\gamma_{10} \simeq 2\pi \times \unit[1.2]{MHz}$).
Consequently, the incoming field resides in the linear dispersive region, where all the spectral components are subject to the same group delay.

In the case where $\sigma$ is $\unit[105]{ns}$ [see \figpanel{fig:swp_sigma_delay_time}{a}], two distinct sharp peaks emerge in the output.
This phenomenon occurs because the signal bandwidth ($\simeq\unit[10]{MHz}$) for $\sigma = \unit[105]{ns}$ is wider than the linewidth of the atom ($\gamma_{10}$).
Consequently, the output signal experiences distortions caused by the non-homogeneous reflection magnitude and group delay.
The interference between the input wave and the atom emission, characterized by opposite phases, gives rise to the observed double sharp peaks.


\section{Group delay time as a function of probe-frequency detuning}

\begin{figure}[t!]
\includegraphics[width=\linewidth]{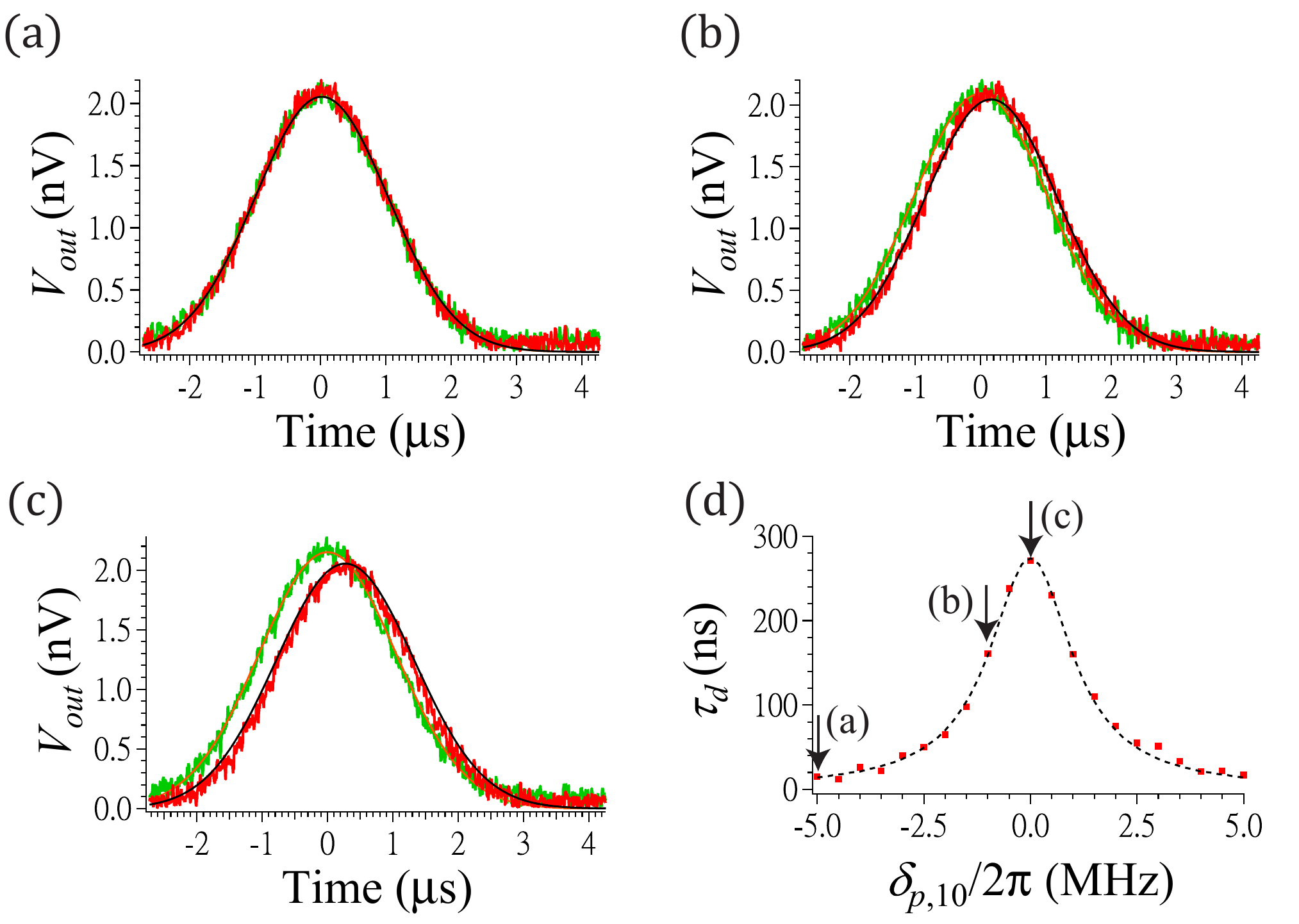}
\caption{
Pulse-envelope dependence on the probe frequency of weak Gaussian pulses with $\sigma \simeq \unit[1]{\upmu s}$.
The probe frequency of the Gaussian pulse is detuned away from the atom by $\delta_{p,10} / 2\pi$ = (a) $\unit[-5]{MHz}$, (b) $\unit[-1]{MHz}$, and (c) $\unit[0]{MHz}$.
Experimental data are shown as red dots (with the atom) and green dots (without the atom, which then is far detuned).
The orange curves are based on Gaussian fitting, and the black curves are based on optical Bloch equations and the input-output relation~\cite{Lin2022} with the parameters of Device 2 in \tabref{tab:qubit_param}.
The extracted $\tau_d$ is (a) \unit[15]{ns}, (b) \unit[161]{ns}, and (c) \unit[271]{ns}.
(d) Summary of extracted $\tau_d$ as a function of $\delta_{p,10}/2\pi$.
The red dots are extracted data and the dashed curve is a simulation.
}
\label{fig:time_domain_detune_pro_freq}
\end{figure}

In this section, for Device 2 with a fixed $\omega_{10} = \unit[4761.62]{MHz}$, the probe frequency $\omega_p / 2\pi$ of the Gaussian pulses is swept from $\omega_{10} / 2\pi - \unit[5]{MHz}$ to $\omega_{10} / 2\pi + \unit[5]{MHz}$, where $\sigma \simeq \unit[1]{\upmu s}$.
The simulation procedure follows the same steps as described in \secref{sec:delayVSsigma}.
In \figref{fig:time_domain_detune_pro_freq}, the simulated black curves show good agreement with the measured data (red dots).
The values of the group delay time $\tau_d$ obtained from the simulation in \figpanel{fig:time_domain_detune_pro_freq}{d} are consistent with the results from the frequency-domain measurements in \figpanel{fig:s2_power_dep}{f}, but display reduced fluctuations.  
The maximum $\tau_d$ occurs when the probe frequency of the Gaussian pulse is on resonance with the atom.
This corresponds to the most significant change in phase response, as shown in \figpanel{fig:s2_power_dep}{c}.

In \figpanel{fig:time_domain_detune_pro_freq}{b}, when the detuning $\delta_{p,10} / 2\pi = \unit[-1]{MHz}$, $\tau_d$ decreases from its maximum \unit[271]{ns} (on resonance, $\delta_{p,10} = 0$) to \unit[161]{ns}.
In \figpanel{fig:time_domain_detune_pro_freq}{a}, where the detuning is increased to $\delta_{p,10} / 2\pi = \unit[-5]{MHz}$, $\tau_d$ decreases further to \unit[15]{ns}.
Therefore, by detuning the probe frequency of the Gaussian pulses, we can adjust $\tau_d$ in the range of 0 to \unit[271]{ns} to control the effect of positive group delay.
This approach complements two methods presented in the main text, enhancing our ability to manipulate the group delay of the light.


\section{Pulse-envelope evolution as a function of probe power}
\label{sec:delayVSprobe_power}

\begin{figure}[t!]
\includegraphics[width=\linewidth]{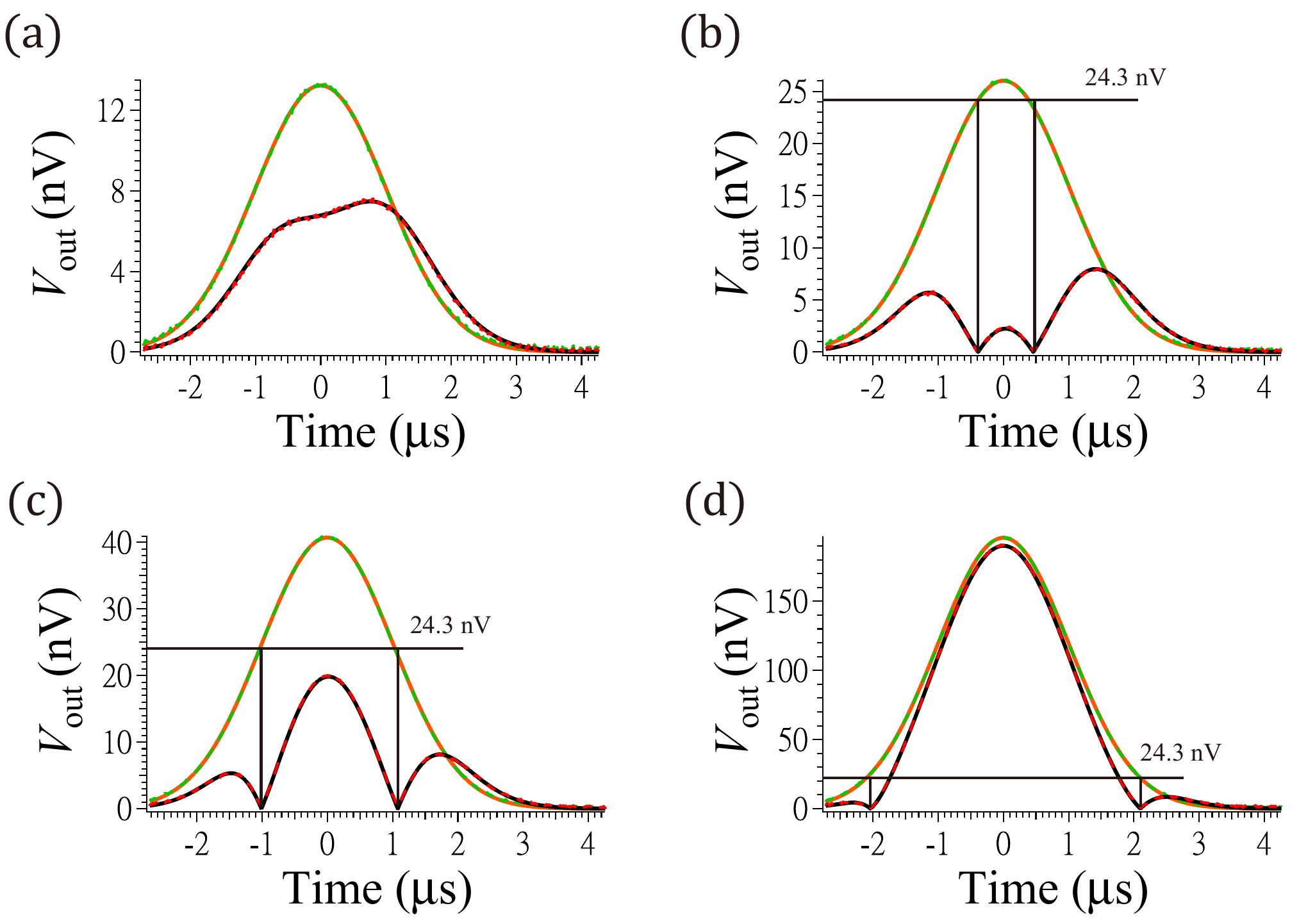}
\caption{
Pulse-envelope variation for different probe powers ($P_p$) of the Gaussian pulse with $\sigma \simeq \unit[1]{\upmu s}$: (a) $\unit[-147.7]{dBm}$, (b) $\unit[-141.7]{dBm}$, (c) $\unit[-137.7]{dBm}$, and (d) $\unit[-123.7]{dBm}$.
Experimental data are shown in red dashed curves (with the atom) and green dashed curves (without the atom, which then is far detuned).
The orange curves are based on Gaussian fitting and the black curves are based on optical Bloch equations and the input-output relation~\cite{Lin2022} with the parameters of Device 2 in \tabref{tab:qubit_param}.
}
\label{fig:time_domain_pro_power}
\end{figure}

In this section, we investigate the behavior of Device 2 by sending Gaussian pulses with different probe powers, while keeping the detuning $\delta_{p,10} = 0$ and the pulse width $\sigma \simeq \unit[1]{\upmu s}$ constant.
As discussed in \secref{sec:delayVSsigma}, for this pulse width the bandwidth of the incoming Gaussian pulses ($\simeq \unit[1]{MHz}$) falls within the linewidth of the artificial atom ($\gamma_{10}$).
Hence, in the time domain, where $\sigma \simeq \unit[1]{\upmu s}$ is much larger than the atom's response time $T_2 = 1 / \gamma_{10} \simeq \unit[135]{ns}$, the atom sees the input approximately as a continuous wave.
With this approximation, we can estimate the output response by utilizing the reflection coefficient given in \eqref{eq:r}. 
In \figref{fig:time_domain_pro_power}, we observe that the time-domain results indeed exhibit the same trend as the power dependence of $|r|$ at $\delta_{p,10} = 0$ shown in \figpanel{fig:s2_power_dep}{e}.

The cases depicted in \figref{fig:time_domain_pro_power} can be divided into two regions based on the input voltage of $\unit[24.3]{nV}$, which corresponds to the singularity point $|r| = 0$ at $P_p = \unit[-142.3]{dBm}$ in \figpanel{fig:s2_power_dep}{e}.
The input voltage is equivalent to the far-detuned curves in \figref{fig:time_domain_pro_power} (orange curves).
In \figpanel{fig:time_domain_pro_power}{a}, when the peak input voltage is smaller than \unit[24.3]{nV}, the input pulse undergoes distortion in both magnitude and phase due to the power dependence of $r$.

As depicted in \figpanels{fig:time_domain_pro_power}{b}{d}, when the peak input voltage exceeds \unit[24.3]{nV}, indicated by horizontal lines, the amplitude around the peak corresponds to $\re{[r]} > 0$. 
When the input voltage $< \unit[24.3]{nV}$, the other amplitude corresponds to $\re{[r]} < 0$.
This leads to the emergence of two dips in the output waveform, indicating that the input amplitudes have reached the singularity point. 
Moreover, with a further increase in the probe power, the atom becomes saturated, causing the output waveform to approach the input waveform, corresponding to $r$ reaching unity.


\section{Autler--Townes splitting as a function of control power}

\begin{figure}[t!]
\includegraphics[width=\linewidth]{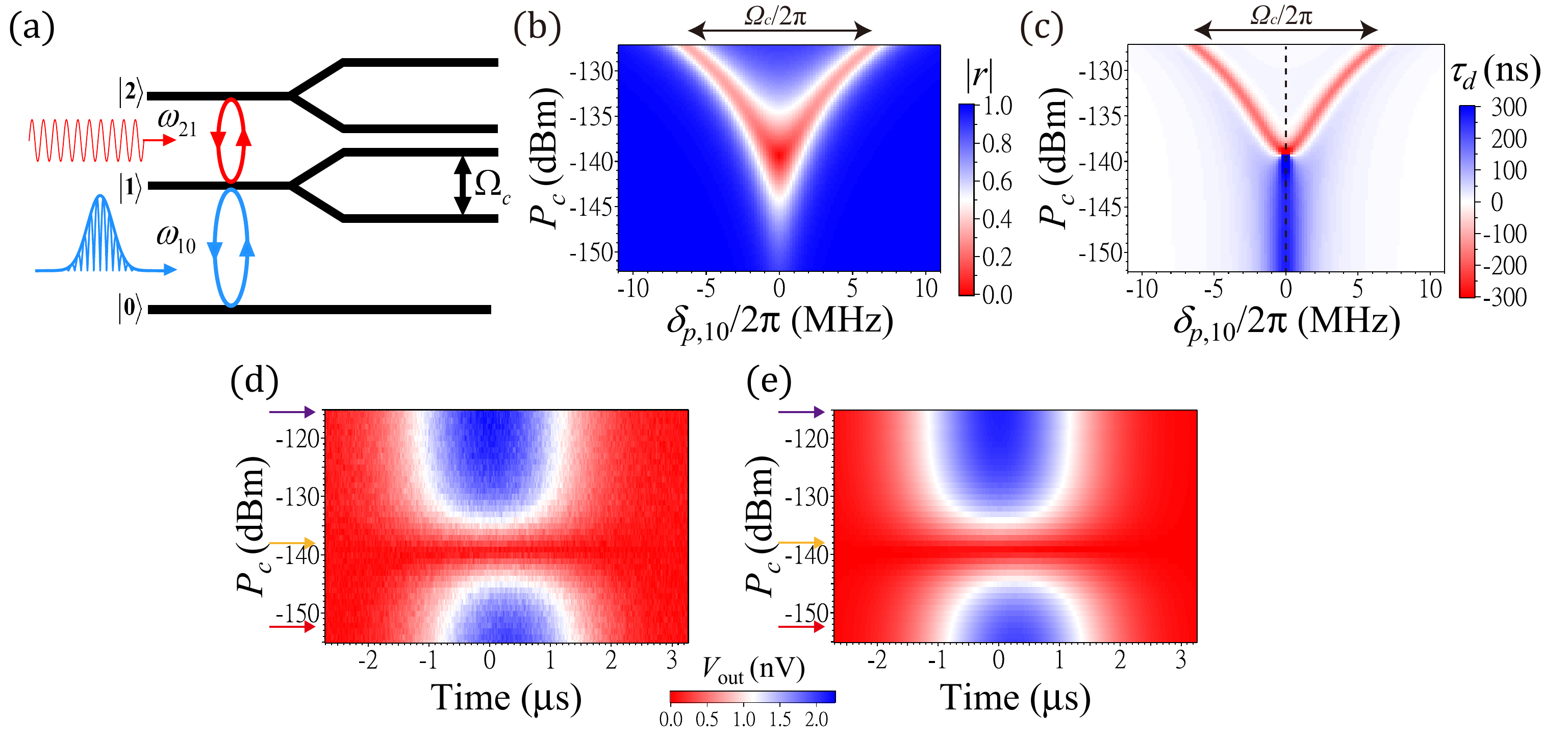}
\caption{
(a) Dressed 3-level system diagram showing the Autler--Townes splitting.
A strong control tone dresses the $\ket{1} \leftrightarrow \ket{2}$ transition of the artificial atom, splitting the levels $\ket{1}$ and $\ket{2}$ individually into dressed states with a splitting of $\Omega_c$ for each.
(b) Theoretical simulation of $|r|$ [corresponds Fig.~3(a) in the main text], illustrating the Autler--Townes splitting of the artificial atom, based on \eqref{eq:r2} with the parameters in \tabref{tab:qubit_param} for Device 2.
(c) The theoretical group delay time $\tau_d$ [corresponds to Fig.~3(b) in the main text] as a function of $\delta_{p,10}$ and $P_c$. When $\delta_{p,10} = 0$, the theoretical $\tau_d$ along the dashed line corresponds to the solid deep blue curve in Fig.~3(e) in the main text.
(d) Measured $V_{\rm out}$ of the Gaussian-pulse ($\sigma \simeq \unit[1]{\upmu s}$) probe field as a function of time and the control-tone power $P_c$. The extracted $\tau_d$ corresponds to the pink data points in Fig.~3(e) in the main text.
(e) The corresponding theoretical prediction for (d) is simulated based on \eqsref{eq:OBE_3Lv_rho10}{eq:OBE_3Lv_rho21} and the input-output relation~\cite{Lin2022} with the parameters in \tabref{tab:qubit_param} for Device 2. 
The arrows in (d) and (e) indicate line cuts in Fig.~3(f) in the main text.
}
\label{fig:ATS}
\end{figure}

In this section, we employ two-tone spectroscopy to investigate our system in Device 2.
On top of the sweeping weak probe tone (approximately $\unit[-162.3]{dBm}$), we apply a control tone that is resonant with the $\ket{1} \leftrightarrow \ket{2}$ transition of the atom.
By utilizing the Autler--Townes splitting (ATS) effect depicted in \figpanel{fig:ATS}{a}, we demonstrate effective switching on and off of the group delay of the light.
This allows us to explore fine-tuning the system from positive to negative group delay for the light.
In \figpanel{fig:ATS}{b}, as the power of the control tone ($P_c$) increases, the $\ket{0} \leftrightarrow \ket{1}$ transition gradually splits into two transitions from the ground state to dressed states due to ATS. 

The reflection coefficient for the weak probe in this case is given by~\cite{hoi2013thesis}
\be
r = 1 - \frac{2 \Gamma_{10}}{2 \mleft( \gamma_{10} + i \delta_{p,10} \mright) + \Omega_c^2 / \mleft[ 2 \gamma_{20} + 2 i \mleft( \delta_{p,10} + \delta_{c,21} \mright) \mright]} ,
\label{eq:r2}
\ee
were $\delta_{c,21} = \omega_c - \omega_{21}$ is the detuning of the control-tone frequency $\omega_c$ from the $\ket{1} \leftrightarrow \ket{2}$ transition frequency $ \omega_{21}$ and the Rabi frequency $\Omega_c = k_{21} \sqrt{P_c}$ of the control tone is proportional to the square root of the control-tone power.
We can extract $\gamma_{20}$ from experimental data [Fig.~3(a) in the main text] using \eqref{eq:r2}, utilizing the parameters $\gamma_{10}$, $\Gamma_{10}$, and $k_{10}$) from \tabref{tab:qubit_param} (Device 2), and the coupling-constant relationship between $k_{21}$ and $k_{10}$.
Given that $\Gamma_{21}$ is estimated to be twice $\Gamma_{10}$~\cite{Koch2007}, and $k_{10}$ is proportional to the square root of $\Gamma_{10}$~\cite{cheng2023tuning}, we can approximate $k_{21} \simeq \sqrt{2} k_{10}$. 
The value of $\gamma_{20}$ is listed in \tabref{tab:qubit_param}.
Consequently, we plot the theoretical ATS spectrum based on \eqref{eq:r2}, illustrated in \figpanel{fig:ATS}{b}.
All the fitting is performed without any free parameters. 
The results demonstrate excellent agreement between theoretical simulations and experimental results. 

In the time domain, we send Gaussian pulses with $\sigma \simeq \unit[1]{\upmu s}$ to the system, while simultaneously applying a continuous control tone with varying power levels. 
The output Gaussian pulses for different control power levels are shown in \figpanel{fig:ATS}{d}, which matches well with simulation results in \figpanel{fig:ATS}{e}.
Notably, the negative-group-delay light is observed in the power range of $\unit[-139.24]{dBm}$ to $\unit[-136.21]{dBm}$, as shown in Fig.~3(e) in the main text.

For the simulations in the time domain, we assume that the atom starts in its ground state.
The time evolution of the elements of the density matrix for the three-level atom is then given by~\cite{hoi2013thesis}
\begin{flalign}
    \partial_t \rho_{10} &= \mleft( - i \delta_{p,10} - \gamma_{10} \mright) \rho_{10} + \frac{i}{2} \Omega_c(t) \rho_{20} + \frac{i}{2} \Omega_p(t) \rho_{00} ,
    \label{eq:OBE_3Lv_rho10}
    \\
    \partial_t \rho_{20} &= \frac{i}{2} \Omega_c(t) \rho_{10} - \mleft[ i \mleft( \delta_{p,10} + \delta_{c,21} \mright) + \gamma_{20} \mright] \rho_{20} - \frac{i}{2} \Omega_p(t) \rho_{21} ,
    \label{eq:OBE_3Lv_rho20}
    \\
    \partial_t \rho_{21} &= - \frac{i}{2} \Omega_p(t) \rho_{20} - \mleft( i \delta_{c,21} + \gamma_{21} \mright) \rho_{21} .
    \label{eq:OBE_3Lv_rho21}
\end{flalign}
Combined with the input-output relation~\cite{Lin2022}, the positive- and negative-group-delay time dynamics of the output response can be numerically solved, as shown in \figpanel{fig:ATS}{e}.


\section{Group delay and reflection coefficient in the weakly probed atom-waveguide system}
\label{sec:GroupDelayAndReflection}

In this section, we begin by deriving the output response of the atom-mirror system in the frequency domain for a given input signal.
Then, we use the narrowband property of our input signal to simplify the output expression and derive the expression for the group delay time utilized in this work. 

An artificial atom is typically a nonlinear system characterized by its power dependence.
The emission of the artificial atom is determined by its state, which naturally decays over time, resulting in a time-variant system.
To use the atom as a filter in our work, we consider the following schemes: 
firstly, by utilizing the weak probe, the system experiences a weak excitation.
This results in the atom predominantly remaining in its ground state and linearizes the system response.
Secondly, the use of a dilution refrigerator helps to maintain the system in a stable ground state with negligible thermal fluctuations until the input probe pulse arrives.
Under these conditions, the system remains linear and time-invariant throughout the entire duration of the experiment.

Based on the power dependence of the spectroscopy~\figpanel{fig:s2_power_dep}{e}, we know that the atom exhibits a linear response at low probe power.
To analyze this linear region, we apply perturbation theory~\cite{Bender1999} to derive an analytical solution.
Specifically, the small parameter used for determining the order of perturbation is $\Omega_p / \gamma_{10}$.
We expand the system density matrix to the first order:
\be
\op{\rho}(t) = \op{\rho}^{(0)}(t) + \op{\rho}^{(1)}(t).
\label{Eq:rho_3Lv}
\ee
Thorough this section, the superscript indicates the order of perturbation.

We consider a three-level Hamiltonian with corresponding dissipator~\cite{Kumar2016}
\begin{flalign}
\op{H} = \frac{\hbar}{2}
\begin{pmatrix}
 	0 & \Omega_p(t) & 0 \\
 	\Omega_p(t) & - 2 \delta_{p,10} & \Omega_c \\
 	0 & \Omega_c & - 2 \mleft( \delta_{p,10} + \delta_{c,21} \mright) \\
\end{pmatrix},
\label{eq:H_3Lv}
\\
\lind{\op{\rho}} = 
\begin{pmatrix}
 	\Gamma_{10} \rho_{11} & - \gamma_{10} \rho_{01} & - \gamma_{20} \rho_{02} \\
 	- \gamma_{10} \rho_{10} & \Gamma_{21} \rho_{22} - \Gamma_{10} \rho_{11} & - \gamma_{21} \rho_{12} \\
 	- \gamma_{20} \rho_{20} & - \gamma_{21} \rho_{21} & - \Gamma_{21} \rho_{22} \\
\end{pmatrix}.
\label{eq:L_3Lv}
\end{flalign}

We then solve the master equation
\be
\frac{d \hat\rho}{dt} = - \frac{i}{\hbar} \comm{\hat\rho}{\hat H} + \lind{\hat\rho}.
\label{Eq:Master_eq_3Lv}
\ee
by employing the Laplace transform with the initial condition that the atom is in its ground state.
Iteratively solving each order of perturbation, we acquire the first-order expansion, which is expressed as
\be
\op{\rho}(s) = 
\begin{pmatrix}
	1 & \rho_{10}^{(1)}(s)^* & \rho_{20}^{(1)}(s)^* \\
	\rho_{10}^{(1)}(s) & 0 & 0 \\
	\rho_{20}^{(1)}(s) & 0 & 0 \\
\end{pmatrix},
\ee
where
\begin{flalign}
\rho_{10}^{(1)}(s) = \frac{i}{2} \Omega_p(s) \frac{s + i \mleft( \delta_{p,10} + \delta_{c,21} \mright) + \gamma _{20}}{\mleft( s + i \delta_{p,10} + \gamma_{10} \mright) \mleft[ s + i \mleft( \delta_{p,10} + \delta _{c,21} \mright) + \gamma _{20} \mright] + \frac{\Omega_c^2}{4}} ,
\label{Eq:rho10_s}
\\
\rho_{20}^{(1)}(s) = \frac{i}{2} \Omega_p(s) \frac{i \Omega _c / 2}{\mleft( s + i \delta_{p,10} + \gamma_{10} \mright) \mleft[ s + i \mleft( \delta_{p,10} + \delta _{c,21} \mright) + \gamma_{20} \mright] + \frac{\Omega _c^2}{4}} .
\end{flalign}
Here, we denote by $f(s) = \laplace{f(t)}$ the Laplace transform of a time-varying function $f(t)$ throughout this section.
Further investigation of the second-order expansion allows us to obtain non-zero corrections for coherence and population.
The first-order expansion allows us to approximate the atom to be predominantly in its ground state, exhibiting time invariance.

The input-output relation for the probe tone~\cite{Lin2022} after Laplace transformation can be expressed as
\be
\Omega_{\rm out}(s) = \Omega_p(s) + 2 i \Gamma_{10} \rho_{10}(s),
\label{Eq:IO_s}
\ee
where $\Omega_{\rm out}$ is the coherent output voltage in Rabi-frequency scale.
Plugging \eqref{Eq:rho10_s} into \eqref{Eq:IO_s}, we obtain the transfer function
\be
\begin{split}
T(s) = \frac{\Omega _{\rm out}(s)}{\Omega _p(s)} = 1 - \frac{\Gamma_{10}}{s + i \delta_{p,10} + \gamma_{10} + \frac{\Omega_c^2}{4 \mleft[ s + i \mleft( \delta_{p,10} + \delta_{c,21} \mright) + \gamma_{20} \mright]}} .
\end{split}
\label{Eq:TF_s}
\ee
The independence of \eqref{Eq:TF_s} from $\Omega_p(t)$ shows the linearity of the system.
Here we assign $s = i \omega_e$ to investigate the frequency response, where $\omega_e$ represents the relative distance from the carrier frequency $\omega_p$.
The reflection coefficient $r$ used in the spectroscopy, obtained from solving the steady-state solution of \eqref{Eq:Master_eq_3Lv} as $\Omega_p(t) = \Omega_p \ll \gamma_{10}$, corresponds to the case $r = T(0)$ and is given in \eqref{eq:r2}.

Given an input pulse $\Omega_p(t) = \Omega_p \exp \mleft[ - \mleft( t - t_0 \mright)^2 / \mleft( 2 \sigma^2 \mright) \mright]$, where the pulse arrival time $t_0$ is much later than the origin $t = 0$ (i.e., $t_0 \gg \sigma$), we can neglect the truncation effect in $\Omega_p(s)$.
This approximation allows us to treat $\Omega_p(s)$ as the Fourier transform of $\Omega_p(t)$, which is also a Gaussian. 
The spectral width of $\Omega_p(s)$, which is $\sigma^{-1}$ for a Gaussian pulse, is assumed to be much narrower than both the carrier frequency $\omega_p$ and the atom linewidth $\gamma_{10}$.
Under this assumption, the entire envelope can be approximated as evolving with the nominal carrier frequency $\omega_p$, which corresponds to $\delta_{p,10} = \omega_p - \omega_{10}$ in \eqref{eq:r2}.

Inside $\Omega_p(s)$, the phase of each spectral component of the envelope is given by $\phase{\laplace{\Omega_p(t)}}(\omega_e) = \omega _e t + \phi_{\omega_e}$, where $\phi_{\omega_e}$ is a constant in time and $\omega_e$ denotes the frequency of the envelope spectral component.
This phase is subject to a phase shift
\be
\phase T(\omega_e) \approx \phase T(0) + \omega_e \mleft. \frac{\partial \phase{T}}{\partial \omega_e} \mright|_{s = 0}.
\label{Eq:TF_phase}
\ee
Defining the group delay~\cite{Oppenheim1997}
\be
\tau_d = - \frac{\partial \phase{T}}{\partial \omega_e} ,
\label{Eq:gd_H_def}
\ee
we see that the negative sign in \eqref{Eq:gd_H_def} leads to a positive $\tau_d$ when the slope in \eqref{Eq:TF_phase} is negative.
Using \eqsref{Eq:TF_s}{Eq:gd_H_def}, the output phase is
\be
\begin{split}
\phase{\laplace{\Omega_{\rm out}(t)}}(\omega _e) &= \phase{T}(\omega_e) + \phase{\laplace{\Omega_p(t)}}(\omega_e) \\
&\approx \omega_e(t - \tau_d) + \phi_{\omega _e} + \phase{T}(0) \\
&= \phase{\laplace{\Omega_p(t - \tau _d)}}(\omega_e) + \phase{r} .
\end{split}
\label{Eq:out_phase}
\ee

From \eqref{Eq:out_phase} we can see that the output envelope is delayed by a time $\tau_g$ compared to the input from the phase perspective.
Using the change of variable $\omega = \omega_e + \omega_p$, we can rewrite \eqref{Eq:gd_H_def} as
\be
\begin{split}
	\tau_d &= -\frac{\partial \phase{T(0)}}{\partial \omega} \\
        &= - \frac{\partial \phase{r}}{\partial \omega} .
\end{split}
\label{Eq:gd_r_def}
\ee
Similarly, for the magnitude dispersion:
\be
\begin{split}
|T(i \omega_e)| &\approx |T(0)| + \omega_e \mleft. \frac{\partial |T|}{\partial \omega_e} \mright |_{s = 0} \\
&= |r|.
\end{split}
\label{Eq:TF_mag}
\ee
In \eqref{Eq:TF_mag}, the expression $\mleft. \frac{\partial |T|}{\partial \omega_e} \mright |_{s = 0} = \frac{\partial |r|}{\partial \delta_{p,10}}$ is complicated, but we can observe that the slope is zero at $\delta_{p,10} = 0$ from the fitting curve in \figpanel{fig:s2_power_dep}{b}, resulting in the only contribution $|T(0)| = |r|$.
When $\delta_{p,10} \ne 0$, the magnitude error can be minimized by selecting a larger $\sigma$, which has a narrower spectrum width, as shown by the results in \secref{sec:delayVSsigma}.

Summarizing from \eqref{Eq:out_phase} to \eqref{Eq:TF_mag}, the output response is
\be
\begin{split}
\Omega_{\rm out}(t) &= \invlaplace{T(s) \laplace{\Omega_p(t)}} \\
&\approx \invlaplace{r \laplace{\Omega_p(t - \tau_d)}} \\
&= r \invlaplace{\laplace{\Omega_p(t - \tau_d)}} \\
&= r \Omega_p(t - \tau_d).
\end{split}
\label{Eq:output_signal}
\ee
Based on \eqref{Eq:output_signal}, the output corresponds to the delayed/advanced version of the input and is additionally rescaled by $r$, given that the spectral content of the envelope is concentrated near the carrier frequency.
This provides a straightforward interpretation of our time-domain results and establishes a direct connection to the spectroscopy results.
Furthermore, this interpretation holds true for our weakly probed two- or three-level systems because the phenomenon of positive/negative group delay for light is a fundamental characteristic of linear systems. 
The expressions provided in Eqs.~(\ref{Eq:gd_r_def}) and (\ref{Eq:output_signal}) are used in Eq.~(2) and Fig.~1 of the main text, respectively.


\section{Deriving effective two-level system reflection coefficient for Device 2}

In this section, we derive the effective reflection coefficient of our three-level-atom case for Device 2.
Starting from \eqref{eq:r2}, we make a Taylor expansion of the denominator in \eqref{eq:r2} near $\delta_{p,10} = 0$, with $\delta_{p,10} / \gamma_{10} \ll 1$ the small parameter ($\delta_{p,10} / \gamma_{20}$ is also small because $\gamma_{20} > \gamma_{10}$).
Equation~(\ref{eq:r2}) can then be expanded to
\be
r \approx 1 - \frac{\Gamma_{10}}{1 - \mleft( \frac{\Omega_c}{2 \gamma_{20}} \mright)^2} \frac{1}{\gamma_{10} \frac{1 + \frac{\Omega_c^2}{4 \gamma_{10} \gamma_{20}}}{1 - \mleft( \frac{\Omega_c}{2 \gamma_{20}}\mright)^2} + i \delta_{p,10}} .
\label{Eq:TF_s_approx}
\ee
By direct comparison to the reflection coefficient for the case of the two-level system, we define the effective rates
\begin{flalign}
\Gamma &= \frac{\Gamma_{10}}{1 - \mleft( \frac{\Omega_c}{2 \gamma_{20}} \mright)^2} , \\
\gamma &= \gamma_{10} \frac{1 + \frac{\Omega_c^2}{4 \gamma_{10} \gamma_{20}}}{1 - \mleft( \frac{\Omega_c}{2 \gamma_{20}} \mright)^2} , \\
\Gamma^n &= \gamma - \frac{\Gamma}{2} = \Gamma^n_{10} \frac{1 + \frac{\Omega_c^2}{4 \Gamma^n_{10} \gamma_{20}}}{1 - \mleft( \frac{\Omega_c}{2 \gamma_{20}} \mright)^2} .
\end{flalign}
We then immediately arrive at the effective two-level reflection coefficient, which is expressed as
\be
r = 1 - \frac{\Gamma}{\gamma + i \delta_{p,10}} .
\label{Eq:eff_2LV}
\ee
This expression is used as Eq.~(1) in the main text.

For the group-delay calculation, we can plug \eqref{Eq:eff_2LV} into \eqref{Eq:gd_r_def}, yielding
\be
\tau_d = \frac{\frac{\Gamma}{\gamma}}{1 + \mleft( \frac{\delta_{p,10}}{\gamma} \mright)^2} \frac{\frac{\Gamma}{2} - \mleft( \Gamma^n - \frac{\delta_{p,10}^2}{\gamma} \mright)}{\mleft( \frac{\Gamma}{2} - \Gamma^n \mright)^2 + \delta_{p,10}^2} .
\label{eq:eff_2LV_tau_d}
\ee
This expression is applicable to the region where $\delta_{p,10} \approx 0$ in the case of a three-level system for Device 2, and it is valid for the entire spectrum of a two-level system as long as $\Omega_c = 0$.
The expression for $\tau_d$ used as Eq.~(2) in the main text is derived by setting $\delta_{p,10} = 0$ in \eqref{eq:eff_2LV_tau_d}.
For the case of the two-level system, according to \eqref{eq:eff_2LV_tau_d}, when twice the magnitude of $|\delta _{p,10}| = \sqrt{\gamma_{10} \mleft( \Gamma_{10}^n - \Gamma_{10} / 2 \mright)}$ at $\tau_d = 0$, it separates the negative-group-delay and positive-group-delay regions.
The second term in the numerator of \eqref{eq:eff_2LV_tau_d} suggests that the detuning $\delta_{p,10}$ can attenuate the effect of the non-radiative decay rate, resulting in the generation of an off-resonance positive-group-delay region in the section related to radiative decay tuning as discussed in the main text in Fig.~2(d) (orange).


\bibliography{supplement}